\def\journal#1, #2, #3, #4 { {\sl #1~}{\bf #2~} (#3)  #4 }

\def\prl{\journal Phys. Rev. Lett., }

\def\cmp{\journal Comm. Math. Phys., }

\def\np{\journal Nucl. Phys., }

\def\pl{\journal Phys. Lett., }

\def\ijmp{\journal Int. J. Mod. Phys., }
\catcode`\@=11
\def\marginnote#1{}
\newcount\hour
\newcount\minute
\newtoks\amorpm
\hour=\time\divide\hour by60
\minute=\time{\multiply\hour by60 \global\advance\minute
by-\hour}\edef\standardtime{{\ifnum\hour<12
\global\amorpm={am}%
        \else\global\amorpm={pm}\advance\hour by-12 \fi
        \ifnum\hour=0 \hour=12 \fi
        \number\hour:\ifnum\minute<10
0\fi\number\minute\the\amorpm}}
\edef\militarytime{\number\hour:\ifnum\minute<10
0\fi\number\minute}

\def\draftlabel#1{{\@bsphack\if@filesw {\let\thepage\relax
   \xdef\@gtempa{\write\@auxout{\string
      \newlabel{#1}{{\@currentlabel}{\thepage}}}}}\@gtempa
   \if@nobreak \ifvmode\nobreak\fi\fi\fi\@esphack}
        \gdef\@eqnlabel{#1}}
\def\@eqnlabel{}
\def\@vacuum{}
\def\draftmarginnote#1{\marginpar{\raggedright\scriptsize\tt#1}}
\def\draft{\oddsidemargin -.5truein
        \def\@oddfoot{\sl preliminary draft \hfil
        \rm\thepage\hfil\sl\today\quad\militarytime}
        \let\@evenfoot\@oddfoot \overfullrule 3pt
        \let\label=\draftlabel
        \let\marginnote=\draftmarginnote

\def\@eqnnum{(\theequation)\rlap{\kern\marginparsep\tt\@eqnlabel}%
\global\let\@eqnlabel\@vacuum}  }

\def\numberbysection{\@addtoreset{equation}{section}
        \def\theequation{\thesection.\arabic{equation}}}

\def\underline#1{\relax\ifmmode\@@underline#1\else
 $\@@underline{\hbox{#1}}$\relax\fi}

\catcode`@=12
\relax
\numberbysection
\def\beq{\begin{equation}}
\def\eeq{\end{equation}}
\def\beqa{\begin{eqnarray}}
\def\eeqa{\end{eqnarray}}
 \def\nnn{\nonumber \\}

\def\hhat{{\widehat h}}
\def\lfloorhat{{\hat \lfloor}}
\def\rfloorhat{{\hat \rfloor}}

\def\demi{{1\over 2}}

\def\Jhat{{\widehat J}}

\def\Je{J^e{}}

\def\Jehat{{\Jhat^e}{}}

\def\mhat{{\widehat m}}
\def\nhat{{\widehat n}}

\def\shat{{\widehat s}}

\def\varpihat{{\widehat \varpi}}

\def\mhat{{\widehat m}}
\def\nhat{{\widehat n}}

\def\Shat{{\widehat S}}

\def\Sb{{\overline  S}}
\def\St{\widetilde S}
\def\Stb{{\overline{\widetilde S}}}
\def\Sbhat{{\widehat {\overline S}}}

\def\zb{{\bar z}}

\def\Vb{{\overline V}}

\def\nub{\bar \nu}

\def\Jgen#1 {  {\underline J_{#1}} }
\def\Kgen#1 {  {\underline K_{#1}} }
\def\Jgenp#1 #2 {(J_{#1}+{#2},\Jhat_{#1})}
\def\Jgenm#1 #2 {(J_{#1}-{#2},\Jhat_{#1})}
\def\Jg#1 {J_{#1},\Jhat_{#1}}
\def\Jgp#1 #2 {J_{#1}+{#2},\Jhat_{#1}}
\def\Mgen#1 {{\underline M_{#1}}}
\def\mgen#1 {{\underline m_{#1}}}

\def\ms{m^\circ{}}

\def\mshat{ {\mhat^\circ}{}}

\def\Vt{{\widetilde V}}

\def\Vb{\overline V}
\def\Vtb{\overline{\Vt}}

\def\fin{\end{document}}

\def\Jgen#1 {  {\underline J_{#1}} }
\def\Jgenp#1 #2 {(J_{#1}+{#2},\Jhat_{#1})}
\def\Jgenm#1 #2 {(J_{#1}-{#2},\Jhat_{#1})}
\def\Jg#1 {J_{#1},\Jhat_{#1}}
\def\Jgp#1 #2 {J_{#1}+{#2},\Jhat_{#1}}
\def\Mgen#1 {{\underline M_{#1}}}

\def\fusV#1,#2,#3,#4,#5,#6 {f_V(
\Jgen{#1} ,
\Jgen{#2} ,
\Jgen{#3} ,
\Jgen{#4} ,
\Jgen{#5} ,
\Jgen{#6} )}

\def\sixjxi#1,#2,#3,#4,#5,#6
{{\left\{\left . \!\! \,^{#1}_{#2}\,^{#3}_{#4}
\right | \!\, ^{#5}_{#6}\right\}}}

\documentstyle[12pt]{article}
\begin{document}
\tolerance 2000
\hbadness 2000
\begin{titlepage}

\nopagebreak \begin{flushright}

CERN - TH/95-295 \\
hep-th/9512125
 \\
December 1995
\end{flushright}

\vglue 3  true cm
\begin{center}
{\large \bf
NEGATIVE SCREENINGS \\
IN LIOUVILLE THEORY}  \\
\vglue 1.5 true cm
{\bf Jens SCHNITTGER}\\
\medskip
{CERN-TH, 1211 Geneva 23, Switzerland.}\\

\end{center}
\vfill
\vglue 1 true cm
\begin{abstract}
\baselineskip .4 true cm
{\footnotesize
\noindent
We demonstrate how negative powers of screenings arise
as a nonperturbative effect within the operator approach
to Liouville theory. This explains the origin of the corresponding
poles
in the exact Liouville  three point function proposed
by Dorn/Otto and $(\hbox{Zamolodchikov})^2$ (DOZZ) and leads to
a consistent  extension of the operator approach to arbitrary integer
numbers of screenings of both types. The general Liouville three
point
function in this setting is computed without any analytic
continuation
procedure, and found to support the DOZZ conjecture.
We point out the importance
of the concept of free field expansions with adjustable monodromies -
recently advocated by Petersen, Rasmussen and Yu - in the present
context,
 and show that it provides a unifying interpretation
for two types of previously constructed local observables.}
\end{abstract}
\vfill
\end{titlepage}
\section{Introduction and Summary}
In spite of remarkable progress in the understanding of quantum
Liouville
theory, the computation of correlation functions has remained a
thorny issue.
Nonrigorous analytic continuation techniques
\cite{GL}\cite{D}\cite{DK}\cite{BK} have provided results
for genus zero and one amplitudes of minimal matter coupled to
gravity which agree with those obtained from the matrix models. The
extension
of these techniques led to an expression for the general Liouville
three point
functions $\langle e^{\alpha_1\Phi} e^{\alpha_2\Phi}e^{\alpha_3\Phi}
\rangle$ at arbitrary values of the central charge \cite{DO}, which
recently
was rediscovered in \cite{ZZ}. However, a rigorous derivation of
these
results is still missing, and the analytic continuation method by its
very nature prevents us from obtaining a true insight into the
dynamical mechanism
which leads to the complicated form of the Liouville correlation
functions.
The basic difficulty one encounters is that though naively, a Coulomb
gas picture for the theory can be obtained by expanding in powers of
the interaction term - which formally is a screening operator
involving
the semiclassical screening charge $\alpha_-$ - , general Liouville
correlators actually require the insertion of  noninteger  powers of
$\alpha_-$ screenings
\footnote{Another way of saying this is that the dependence of
correlators on the cosmological constant is in general nonanalytic.}.
The operator approach to Liouville theory, developed in a long series
of papers \cite{G1}, provides a well-defined framework within which
it should be possible
to address this problem. In \cite{GS3}, general Liouville
exponentials were constructed
as infinite series in Coulomb-gas-like vertex operators, while no
attempt was made yet to determine their correlation functions.
In the present paper, we intend to make a first step towards
a first principle derivation of  Liouville correlators (within the
elliptic sector
usually considered, that is, for "microscopic" operators
\cite{S}\cite{GM}),
 by
considering an important special case. We will analyze the three
point function
in the situation where the total screening charge
$\alpha_s=2\alpha_0-(\alpha_1+
\alpha_2+\alpha_3)$ is given by a linear combination with integer
coefficients $\alpha_s =s_+\alpha_+ +s_-\alpha_-$ of the two
elementary screening charges
$\alpha_+$ and $\alpha_-$ ,
and  the exponent $-2\alpha$ of at least one of the Liouville
exponentials
$e^{\alpha_i\Phi}$ fulfills the same condition.
(The degenerate sector of Liouville theory, studied extensively in
the older
works \cite{GN}\cite{CGR1}, is recovered when these integers are
positive). For irrational
central charge where $\alpha_+$ and $\alpha_-$ are incommensurate,
these
values lie dense in the continuum.
The reason for the restriction to integer screening numbers of both
types
 can be understood within the concept of multivalued free field
expansions with arbitrary monodromy
recently introduced in ref. \cite{PRY} within the context of
correlators
for $SL(2)$ - Kac-Moody based theories. It provides, at the same
time, a
 unifying interpretation for two apparently different kinds of local
observables constructed in ref. \cite{GS3}. The first type involves
only
the semiclassical screening charge $\alpha_-$ and is continously
connected
to the classical Liouville exponentials by a naive classical limit.
On the other hand, the DOZZ three point function is completely
symmetric
in $\alpha_+$ and $\alpha_-$, just like the second kind of local
observables of
ref. \cite{GS3}, whose interpretation had remained unclear. The
explanation
we will propose is that the two types of observables are nothing but
expansions
with different monodromies of the same Liouville exponentials, to be
chosen according to whether the matrix element considered involves
$\alpha_+$ -screenings or not.

In our analysis we will distinguish two basic cases:
If (at least) one of the four
conditions $s_\pm\ge 0$,  $s_\pm <0$,  $s_\pm\le -p_\pm$, or
$s_\pm > -p_\pm$  is fulfilled, where $p_\pm$
 characterize the integer-valued exponent
 $2\alpha=p_+\alpha_+ +p_-\alpha_-$, then the DOZZ three point
function has a pole; otherwise it is
regular as a function of the screening charge $\alpha_s$
\footnote{(in the generic case
when only one of the exponents $2\alpha_i$ is integer-valued)}.
Only the poles at positive $s_-$ and $s_+=0$ are  expected within the
naive
perturbative picture as resulting
from a charge conservation $\delta$ function for the zero mode
integration,
while the occurrence of  poles at $ s_- <0 $ and at $s_+\ne 0$ is
surprising.
The concept of variable monodromy expansions, together with locality
and the complete
symmetry of the conformal structure under exchange of $\alpha_+$
and $\alpha_-$, provides some explanation
 for the latter, but not for the appearance of poles at negative
screenings.
However, we will derive that the series expansions of ref. \cite{GS3}
for the Liouville exponentials, which superficially involve only
positive
powers of screenings, actually simulate the effect of negative
screenings
when evaluated carefully, due to the nontrivial convergence
properties
of the infinite series. Indeed, using a coherent state basis
(eigenstates
of the screening operators) one finds that the zero mode integration
can be
transformed into a contour integral on the unit circle  over the
eigenvalue
 of the screening operator. The residue at the origin corresponds to
a naive ("perturbative") term by term zero mode integration of the
series
representing the Liouville
exponential, while the other residues represent nonperturbative
effects
which break the charge conservation rules of the naive perturbative
picture.
This interesting mechanism  explains the appearance of poles in the
DOZZ
three point function at negative
values of the screening numbers, the singularity
coming simply from the infinite zero mode volume. In the present
analysis,
where we restrict effectively to a discrete spectrum, the zero mode
volume
is treated as finite, and we obtain the residues of the DOZZ result
at the
poles.
The inclusion of negative powers of screenings leads to an extension
of the operator approach
as elaborated in recent works
\cite{GS3}\cite{GS1}\cite{GR}\cite{CGR1}. It is
an important virtue of the Gervais-Neveu quantization scheme that
negative
powers of screenings can be formulated without the need for analytic
continuation. Indeed, positive and negative screening powers are
related to the
two canonically equivalent fields introduced long ago \cite{GN1}. All
observables must be completely symmetric under the exchange of the
two
free fields, which can be associated with a Weyl reflection with
respect
to the underlying $U_q(sl(2))$ symmetry of the theory, and this
serves as an important guiding principle for the nonperturbative
definition of
the Liouville exponentials. We will show that the three point
function
is invariant under this reflection, so that highest weight
 states which it relates
can be consistently identified, as expected from the two to one
nature
of the B\"acklund transformation relating the Liouville to the free
field.
The resulting theory lives exactly on the poles of the DOZZ three
point function and fulfills the crossing symmetry/locality
conditions.
On the other hand, in the cases where the DOZZ three point function
is finite, the sum of the residues vanishes; this is of course again
a
consequence of the different treatment of the zero mode volume.
However, we will point out briefly that in fact it is possible to
obtain the DOZZ result at these points as well by a formal
renormalization of the expansion
coefficients of the Liouville exponentials.
The resulting ``renormalized exponentials'', which are again local,
 turn out to reproduce
as a special case the dressing operators proposed in ref. \cite{G1}
for
the coupling of minimal matter to gravity.

 \section{Liouville exponentials}
We start by briefly recalling the results of ref. \cite{GS3} that
will be needed here. Liouville fields $e^{-J\alpha_-\Phi}$ with
arbitrary spin $J$ and
conformal weight $\Delta_J=-J-{h\over\pi}J(J+1)$
can be written as an infinite series in Coulomb gas vertex operators:
\beq
\exp[-J\alpha_-\Phi(\sigma,\tau)]=\sum_{n=0}^\infty
{T(\varpi)\over T(\varpi+2n-2J)} a_n^{(J)}(\varpi)
V^{(J)}_{-J}(u){\Vb}^{(J)}_{-J}(v)S(u)^n{\Sb}(v)^n
\label{Liouexp}
\eeq
Here $\varpi$ denotes the (rescaled) free field zero mode which is
real
in the elliptic sector considered here, $u=\tau +\sigma$ and
$v=\tau-\sigma$ are light cone variables formed from the usual
cylinder
coordinates $\tau$ and $\sigma$, and the coefficient
$a_n^{(J)}(\varpi)$  is given by
$$
 a_n^{(J)}(\varpi)=
\left ( {\Gamma(1-h/\pi)\over 2\pi}\right)^{2n} {\lfloor -2J\rfloor_n
\over \lfloor n\rfloor!\lfloor\varpi+1\rfloor_n
\lfloor \varpi -2J+n\rfloor_n }
$$
with
\beq
\lfloor x\rfloor :={\sin (hx) \over \sin (h)},\quad \lfloor
a\rfloor_n
:=\prod_{k=0}^{n-1} \lfloor a+k\rfloor
\label{coeff}
\eeq
The effective Planck constant $h$ is related to the central
charge $C$ by\hfill
\beq
h={\pi \over 12}(C-13 -\sqrt{(C-25)(C-1)})
\label{defh}
\eeq
As usual, $\alpha_-$ denotes the semiclassical screening charge,
given
in terms of $h$ by
\beq
\alpha_-=\sqrt{2h\over \pi}.
\eeq
Furthermore, the Coulomb gas vertex operators are defined as
\beq
V_{-J}^{(J)}(u)=
: e^{2J\sqrt{h/2\pi}\vartheta_1}:
\eeq
where $\vartheta_1$ is a chiral free field obeying
\beq
\Bigl [\vartheta'_1(u_1),\vartheta'_1(u_2) \Bigr ]
=2\pi i\,  \delta'(u_1-u_2)
\eeq
(the meaning of the index $1$ will become clear later on) and
\beq
 S(u)= e^{2ih(\varpi+1)}
 \int _0 ^{u} du' V_{1}^{(-1)}(u')+
 \int _u  ^{2\pi } du' V_{1}^{(-1)}(u').
\label{screening}
\eeq
$S(u)$ is a dimension zero primary field, called screening operator
in the sequel.
Similar formulae hold for the right-moving vertex operators.
We note that powers of $S$ are well-defined without
renormalization\footnote{(Up to a certain
maximal power depending on $h$, but the Coulomb gas operators retain
a well-defined meaning for any power by analytic continuation -see
\cite{GS3}).},
and the same is true for the product $V^{(J)}_{-J}(u)S^n(u)$.
The Coulomb gas operators shift the zero mode by
$$
V_{-J}^{(J)} \varpi = (\varpi-2J) V_{-J}^{(J)}
$$
and
\begin{equation}
S \varpi = (\varpi+2) S,
\label{shift}
\end{equation}
Finally, let us turn to the factor ${T(\varpi)\over
T(\varpi+2n-2J)}$.
In ref. \cite{GS2} it was pointed out that the locality conditions
which determine
the coefficients $a_n^{(J)}(\varpi)$  possess an obvious invariance
generated
by the similarity transformations
\beq
\exp{[-J\alpha_-\Phi]} \to
T(\varpi)\exp{[-J\alpha_-\Phi]}T^{-1}(\varpi)
\label{simtrafo}
\eeq
with arbitrary (smooth) functions $T(\varpi)$, which, using Eqs.
(\ref{shift}),
lead to the first factor in Eq.(\ref{Liouexp}). However, $T(\varpi)$
is not truly
arbitrary but strongly constrained by imposing invariance under the
discrete
remnant of the underlying $SL(2)$-symmetry of the theory -see below.
For reasons
discussed in section \ref{monodromy}, the expansion Eq.
(\ref{Liouexp}) should be used
only for matrix elements of the type
$\langle f'_{\varpi_f}|\exp{[-J\alpha_-\Phi]}|f_{\varpi_i}\rangle $,
$f_\varpi$
 denoting  an arbitrary state in the Fock module over the ground
state $|\varpi \rangle$, with integer values of
\beq
s:=J-\demi(\varpi_f-\varpi_i) \ \in {\bf Z}
\label{s}
\eeq
Eqs. (\ref{shift}) tell us that $s$ is the number of screening
charges
($s\alpha_- =\alpha_s$) needed to connect bra and ket in the above
matrix
element. Naively these matrix elements are nonzero only for
nonnegative
$s$ according to Eq. (\ref{Liouexp}), but actually we will see that
the formal infinite series in Eq. (\ref{Liouexp}) effectively
contains also negative screenings. On the other hand, if we consider
the more general matrix
elements with
\beq
s \in {\bf Z} +{\bf Z}{\pi\over h}
\label{gens}
\eeq
with $\pi/h=\alpha_+/\alpha_-$ equalling the ratio of the two
screening charges, then we should use the expansion
\beq
\exp[-J\alpha_- \Phi] = \sum_{n,\nhat=0}^\infty
{T(\varpi)\over T(\varpi+2m_{n,\nhat})}
a_{n,\nhat}^{(J)}(\varpi)
V^{(J)}_{-J}S^n \Shat^\nhat \Vb^{(J)}_{-J} \Sb^n { \Sbhat} ^\nhat
\label{genexp}
\eeq
where $m_{n,\nhat}:=-J+n+\nhat\pi/h$ and
$$
a_{n,\nhat}^{(J)}(\varpi)=a_n^{(J)}(\varpi){\widehat
a}^{(\Jhat)}_\nhat
(\varpihat)\equiv
$$
$$
\left ( {\Gamma(1-h/\pi)\over 2\pi}\right)^{2n}
{\lfloor -2J\rfloor_n \over \lfloor n\rfloor!
\lfloor\varpi+1\rfloor_n
\lfloor\varpi -2J+n\rfloor_n}
$$
\beq
\times
\left({\Gamma(1-\hhat/\pi) \over 2\pi}\right )^{2\nhat}
{\lfloorhat -2\Jhat\rfloorhat_\nhat
\over \lfloorhat \nhat\rfloorhat!\lfloorhat\varpihat+1\rfloorhat
\lfloorhat
\varpihat-2\Jhat+\nhat\rfloorhat}
\label{gencoeff}
\eeq
Here we have put
$ \hhat:={\pi^2\over h},
\ \lfloorhat x \rfloorhat := {\sin (\hhat x)\over
\sin(\hhat)}, \ \Jhat:=J{h\over\pi}$ and $ \varpihat:= \varpi
{h\over\pi}$.
Likewise, $\Shat$ is given by Eq. (\ref{screening}) with ${h}$
replaced
by ${\hhat}$ everywhere.
Eqs. (\ref{genexp}), (\ref{gencoeff}) are manifestly symmetric in
$\alpha_+,\alpha_-$ (or $h,\hhat$) and so the preference of
$\alpha_-$ over $\alpha_+$
on the left-hand side of Eq. (\ref{genexp}) is purely notational.
In using, different from ref. \cite{GS3}, the same designation for
the
local observables of Eq. (\ref{Liouexp}) and Eq. (\ref{genexp}), we
anticipate
the result of the discussion in section \ref{monodromy} which leads
to their
identification.
We note that here and below,
we consider only the case of zero winding number, that is, of equal
values
of left and right zero mode \cite{GS3}. For later convenience, we
introduce another, equivalent way of writing Eqs. (\ref{Liouexp}) or
(\ref{genexp}) \cite{GS3}:
\beq
\exp{[-J\alpha_-\Phi]}=\sum_{{\underline n}=0}^\infty
{T(\varpi)\over
T(\varpi+2m)}
g^x_{J,x+m}{\overline g}^x_{J,x+m}V^{(J)}_m\Vb^{(J)}_m
\label{gform}
\eeq
where $m:=m_{n,\nhat}$ with $\nhat=0$ in the case of Eq.
(\ref{Liouexp}),
and similarly ${\underline n}=n,\nhat$ or ${\underline n}=n$.
Furthermore, $V^{(J)}_m$ denote normalized vertex operators with
$\langle\varpi|V^{(J)}_m|\varpi+2m\rangle =1$, related to the Coulomb
gas
vertex operators by
\beq
V^{(J)}_{-J}S^n\Shat^\nhat=I^{(J)}_m(\varpi)V_m^{(J)}
\label{Vops}
\eeq
where the normalization factors $I^{(J)}_m(\varpi)$ are given
in the appendix.
The coefficients $g^x_{J,x+m}$ with $x:=\demi(\varpi-\varpi_0)\equiv
\demi(\varpi-1-\pi/h)$  appear as coupling constants in the
fusion and braiding algebra of the chiral vertex operators
\cite{CGR1},
\cite{GS3}. Comparing Eq. (\ref{gform}) with Eq. (\ref{Liouexp}) or
Eq. (\ref{genexp}) we have
\beq
g^x_{J,x+m}{\overline g}^x_{J,x+m}={a}_{\underline n}^{(J)}(\varpi)
|I_m^{(J)}(\varpi)|^2.
\label{grel}
\eeq

\section{Nonperturbative evaluation of matrix\break
  elements}
\label{nonpert}

Due to conformal invariance \cite{GN2}, in order to define the
operators
Eq. (\ref{gform}) it is in principle sufficient
to determine the ground state matrix element
\beq
\langle
\varpi_f|\exp{[-J\alpha_-\Phi]}(\tau=\sigma=0)|\varpi_i\rangle
\label{matel}
\eeq
with  the screening number  $s$ subject to Eq. (\ref{gens}) but
otherwise
 arbitrary parameters. This matrix element is closely
connected to the three point function via the formal
 relations
$$
\vert \varpi_i \rangle =F_i \lim_{\tau \to -\infty} (z\bar
z)^{-\Delta_J}
(\exp{[-J_i\Phi(-i\tau,\sigma)]})_{{\underline n}=0}\vert
\varpi_0
\rangle
$$
$$
\langle \varpi_f \vert =F_f \lim_{\tau \to +\infty} (z\bar
z)^{\Delta_J}
\langle -\varpi_0 \vert
(\exp{[-J_f\Phi(-i\tau,\sigma)]})_{{\underline n}=0}
$$
Here, the subscript on the Liouville exponentials denotes the
$n=\nhat=0$ contribution in Eq. (\ref{gform}),
\beq
z:=e^{\tau+i\sigma},\quad\varpi_0 :=1+\pi/h, \quad J_f=
-\demi(\varpi_0+\varpi_f), \quad
J_i= \demi(\varpi_i-\varpi_0)
\label{extstates}
\eeq
and $F_i,F_f$ are normalization factors which will be determined
later.
The prescription Eq. (\ref{extstates}) relates operators to highest
weight states of the same conformal dimension, as usual. However,
in Liouville theory the operator - state correspondence is actually
a subtle issue \cite{P}\cite{S}\cite{GM}, and this is related to
the problems with defining an $SL(2)$ - invariant vacuum in the
present
framework \cite{LS}.\footnote{Actually, the states
$\vert \pm \varpi_0 \rangle, \langle \pm \varpi_0 \vert $ do not
really exist
in the Hilbert space defined by the simultaneous existence of the two
free fields $\vartheta_1$ and $\vartheta_2$  \cite{LS}. However, the
limits
Eq. (\ref{extstates}) do exist if the $\vartheta_1$ representation is
used;
see section \ref{reflectionamp} for further remarks.}
Therefore,  at present Eq. (\ref{extstates})
should be viewed as a formal prescription. In the
degenerate case studied in the older works
 \cite{GN2}\cite{G1}, where the dimensions of the Liouville fields
fall into
Kac's table and the sums in Eq. (\ref{gform}) truncate, it is
unnecessary
to select the ${\underline n}=0$ contributions as all other ones are
 automatically suppressed in the limit $z\to 0$; however this is not
the case
in the present extended framework where divergent terms would appear.

In this section we will  concentrate on the
case where only $\alpha_-$ screenings are present, i.e.  $\nhat=0$,
and $s \in {\bf Z}$,
so that Eq. (\ref{Liouexp}) is relevant. As a first step, let us
explain why naive charge conservation
 cannot be used when evaluating
matrix elements of Eq. (\ref{Liouexp}), in particular the ground
state
matrix element (\ref{matel}).

 \subsection{Reflection invariance and broken charge conservation}
\label{SL2}

One obvious observation is that the three point function relevant for
the
 coupling of minimal matter to gravity is known \cite{D} to require a
negative number of screening operators, while Eq. (\ref{Liouexp})
contains only positive
screenings, and thus a naive evaluation would give a vanishing result
in this context. However, it is also possible to see purely from
internal
consistency considerations that charge conservation cannot hold:
Consider the general classical solution of the Liouville equation,
\beq
\exp{[-\varphi]} = A'^{-1}(u)A^2(u)B'^{-1}(v) \left(1-{B(v)\over
A(u)}\right)^2,
\label{classfield}
\eeq
where $A(u)$ and $B(v)$ are two arbitrary functions.
In ref. \cite{GN1} the important observation was made that after
gauge-fixing the projectively acting SL(2) symmetry of Eq.
(\ref{classfield}),
\beq
A \to {aA+b\over cA +d}, \quad B \to {aB+b\over cB +d}
\label{sl2symm}
\eeq
by demanding that $A$ and $B$ have diagonal monodromy (i.e. that they
be
periodic up to a multiplicative constant, leading to periodic
free fields), there remains the discrete
symmetry\footnote{For general Toda theories, this becomes the
symmetry
of the theory under the Weyl group for the corresponding Lie algebra
\cite{BG}\cite{ANPS}.}
 $ A \to -1/A,\  B\to -1/B$, which leads to the existence of {\it
two}
equivalent free fields $\vartheta_1,\vartheta_2$ instead of just one.
It was shown in \cite{GN1} that this symmetry extends to the quantum
level,
and that there is a quantum canonical transformation
relating $\vartheta_1$ and $\vartheta_2$. Now certainly all
observables
of the theory (i.e. all functionals of the Liouville field) must be
invariant
under the exchange
\beq
\vartheta_1 \leftrightarrow \vartheta_2,
\label{thetaex}
\eeq
 as this exchange just represents a particular $SL(2)$
transformation.
We will call it reflection symmetry in the following.
Due to the relation
\beq
\tilde \varpi =-\varpi
\label{omegatilde}
\eeq
between the zero modes of $\varpi_1$ and $\varpi_2$,
the vertex operators $\Vt^{(J)}_{-J}{\St}^n \Vtb^{(J)}_{-J}
\Stb^n$ formed using $\vartheta_2$ have
zero mode shifts exactly opposite to those formed from $\vartheta_1$
\footnote{We warn the reader familiar with the previous papers
that the $\Vt$ operators here are different from those of \cite{CGR1}
or \cite{GS3}. There should be no confusion as the latter do not
appear
in the present paper.}:
\beq
[\varpi,\Vt^{(J)}_{-J}{\St}^n \Vtb^{(J)}_{-J}{\Stb}^n]
=+2m \Vt^{(J)}_{-J}{\St}^n \Vtb^{(J)}_{-J} {\Stb}^n
\label{tildeshifts}
\eeq
Now as formally the zero mode shifts in Eq. (\ref{Liouexp}) are
given by $\Delta \varpi \equiv -2m =2J, 2J-1, 2J-2,.....$, (similarly
for Eq. (\ref{genexp})) we see that for generic $J$, there is no
overlap
at all
between the shifts predicted using the two representations! Thus we
conclude that charge conservation must be broken. The only
exception is the case when $2J$ is positive integer\footnote{This is
of
course the case
where the Liouville exponentials are in Kac's table, corresponding to
the
well-understood degenerate sector of Liouville theory
\cite{GN}\cite{CGR1}.}, where the sum in Eq. (\ref{Liouexp})
truncates at $n=2J$ and the two sets of shifts indeed coincide. For
this particular case, the requirement
of reflection invariance Eq. (\ref{thetaex}) was shown in
ref. \cite{GS3} to largely fix the function $T(\varpi)$ in Eqs.
(\ref{Liouexp}),
(\ref{genexp}):\footnote{Compared to ref. \cite{GS3}, we have
replaced a factor
$\sqrt{\lfloor\varpi\rfloor}$ by $\lfloor\varpi\rfloor$, making use
of the freedom Eq. (\ref{Tfreedom}) below.}
\beq
T(\varpi)=
{\Gamma(1-\varpi h/\pi) \lfloor \varpi\rfloor
 \over
\Gamma(1+\varpi)}
\label{T}
\eeq
with the freedom of replacing   $T(\varpi)
\to T(\varpi)T_1 (\varpi)$, where
\beq
{ T_1(\varpi)T_1(-\varpi-2(n-J))\over
T_1(\varpi+2(n-J)) T_1(-\varpi)}=1
\label{Tfreedom}
\eeq
for any allowed value of $n$ and $2J$.
 We will use Eq. (\ref{T}) as an ansatz also for general $2J$.
\footnote{
Actually, $T(\varpi)$ should be independent of $J$ in order to really
drop out of the locality analysis. }

Returning to the general case, how does the theory manage to break
charge
conservation? The answer is that the infinite expansions Eq.
(\ref{Liouexp}) and Eq. (\ref{genexp}) have rather subtle convergence
properties
 (as was remarked already in ref. \cite{GS3}), and must be resummed
and regularized in an appropriate
way to become truly well-defined. In particular, within matrix
elements of the Liouville exponentials it is not allowed
to permute the order of zero mode integration and sum over $n$ or
$n,\nhat$,
which would lead to the naive charge conservation rules.
For better readability, let us first sketch the main idea and
then give a more exact technical derivation.

\subsection{Resummation and screening eigenstates}
The basic idea in order to render matrix elements of Eq.
(\ref{Liouexp})
well-defined is to  insert
a particular complete set of intermediate  states, and then to carry
out
the sum over $n$ {\it before}  the factorization sum.  This complete
set will then have to be chosen such that the $n$ sum can be carried
out explicitly; now since Eq. (\ref{Liouexp}) is an expansion in
powers of screenings, it is natural to consider a basis composed of
eigenstates of the latter.
 Thus we are lead to replace the matrix element
(\ref{matel}) by
\beq
\langle \varpi_f|\exp{[-J\alpha_-\Phi]}|\varpi_i\rangle \quad
\rightarrow \quad
\sum_f\int dy\langle \varpi_f|\exp{[-J\alpha_-\Phi]}|y;f\rangle
\langle y;f|\varpi_i\rangle,
\label{insertion}
\eeq
where $|y;f\rangle$ denotes an eigenstate of the screening operator
$S\Sb(\tau=\sigma=0)\equiv S\Sb$, with eigenvalue parametrized by $y$
and $f$ a quantum number describing a basis of the corresponding
eigenspace,
and $\sum_f\int dy |y;f\rangle\langle y;f|={\bf 1}$.
For fixed $y$, every term of the $n$ - sum in Eq. (\ref{Liouexp})
contributes and the result will be given in terms of a $q$ -
hypergeometric
function at $|q|=1$. The latter is well defined in the case where
$2J$
is integer - this is the origin
 of the restriction on one of the spins\footnote{
Due to the symmetry of the three point function under permutation of
the legs,
our result will be valid also when one of the spins $2J_f,2J_i$
 corresponding to
$\varpi_f$ and $\varpi_i$, rather than $2J$, is integer.}
mentioned in the introduction -   and
 can then be integrated over $y$ to give an exact closed
form expression for the matrix element (\ref{matel}). It is
invariant under the reflection symmetry mentioned above.

\smallskip
To make the above statements precise, let us  first define
eigenstates of $S\Sb$
by
\beq
|y;f_{\varpi_1}\rangle := \sum_{n=-\infty}^\infty e^{-iyn}
(S_0\Sb_0)^n |
f_{\varpi_1}
\rangle
\label{eigenstates}
\eeq
In Eq. (\ref{eigenstates}), $S_0$, $\Sb_0$ denote normalized
screening charges with the property
\beq
(S_0\Sb_0)^\dagger = (S_0\Sb_0)^{-1},
\label{unitarity}
\eeq
Here $e^{iy}$ with $y\in [0,2\pi]$ is the eigenvalue, and
$f_{\varpi_1}$ denotes an arbitrary
Fock state over the ground state $|\varpi_1\rangle$. The value of
$\varpi_1$
will be fixed (up to integers) by the matrix element (\ref{matel})
under consideration.
Of course the unitarity property of $S_0\Sb_0$ is needed for the
eigenvalues $e^{iy}$ to be pure phases and hence for the sum in Eq.
(\ref{eigenstates}) to make sense. The latter is completely analogous
to the formation of position eigenstates by a linear superposition of
plane
waves -the standard Fourier representation $\sum_{n=-\infty}^\infty
e^{2\pi in(x-x_0)}$
of the (periodic) position
space $\delta$ function.
The normalized screening operator $S_0\Sb_0$ is related to $S\Sb$
by
\beq
S_0 \Sb_0= {1\over (2\pi\Gamma(1+{h\over\pi}))^2}{h(\varpi+1)\over
\sin h(\varpi+1)}\Gamma(1-\varpi{h\over\pi})
\Gamma(1+(\varpi+2){h\over\pi})
S\Sb
\label{normscreen}
\eeq
To check that Eq. (\ref{unitarity}) is fulfilled with this
definition,
one first observes that due to the hermiticity properties of the two
free
fields in the elliptic sector, $\vartheta_1^\dagger=\vartheta_2$
\cite{GN1},
one has
\beq
(S\Sb)^\dagger = {\St}{\Stb}
\eeq
Thus if we write
\beq
S_0\Sb_0=N_1(\varpi)S\Sb,
\label{N}
\eeq
the unitarity requirement for $S_0\Sb_0$ becomes
\beq
{1\over |N_1(\varpi)|^2}=S{\St}
\Sb{\Stb}
\label{unitaritycond}
\eeq
Now one has to compute $S\St \Sb \Stb$. Since $\St$
 has a zero mode shift opposite of $S$, we can identify it as a
negative
screening operator. Due to the reflection symmetry Eq.
(\ref{thetaex}), positive and negative screenings must enter into the
theory on exactly the same footing, and the latter, according to the
above,  possess a rigorous definition within the Gervais-Neveu
quantization scheme without the need
for any analytic continuation. The product
$S\St$, which can be shown to exist without
renormalization\footnote{This can be understood as a consequence of
the fact
 that the singular part of $\vartheta_1(u)
\vartheta_2(u')$ is the same as that of
$\vartheta_1(u)\vartheta_1(u')$
\cite{GN1}.}, is a dimension zero
operator with vanishing zero mode shift. As these two data
characterize
a primary field in the free field Hilbert space up to an
$\varpi$-dependent
normalization \cite{GN2}, we conclude that
\beq
S\St  =K_1(\varpi){\bf 1}
\label{SStilde}
\eeq
and thus Eq. (\ref{unitaritycond}) makes sense.
As classically $S(u)\sim A(u)$ \cite{GS1}, the classical limit of Eq.
(\ref{SStilde}) is simply the statement $A\cdot(-1/A)=-1$.
The factor $K_1(\varpi)$ is now easily computed from the
associativity
of the operator product, applied to the expression
$V^{(\demi)}_{-\demi}
(\sigma) S(\sigma) {\St}(\sigma)$. According to Eq. (\ref{Vops}) we
have
\beq
V^{(\demi)}_{-\demi}S=
I^{(\demi)}_\demi(\varpi)V^{(\demi)}_{+\demi}=
I^{(\demi)}_\demi(\varpi)\widetilde{V^{(\demi)}_{-\demi}}
\eeq
where the last equality follows again from Eq. (\ref{tildeshifts}).
Now $\widetilde{V^{(\demi)}_{-\demi}}\tilde
S=\widetilde{I^{(\demi)}_{\demi}
(\varpi)}\widetilde{V^{(\demi)}_{\demi}}$ and thus, using Eq.
(\ref{omegatilde}),
\beq
V^{(\demi)}_{-\demi}(\sigma) S(\sigma) {\St}(\sigma)=
I^{(\demi)}_{\demi}(\varpi)I^{(\demi)}_{\demi}(-\varpi)V^{(\demi)}
_{-\demi}(\sigma)
\eeq
from which we obtain immediately
\beq
K_1(\varpi)=I^{(\demi)}_{\demi}(\varpi+1)
I^{(\demi)}_{\demi}(-\varpi-1)
\label{K1}
\eeq
Explicit computation now shows that $1/K_1(\varpi)$ agrees with
$N_1(\varpi)$
 given by Eqs. (\ref{N}),(\ref{normscreen}) and therefore
Eq. (\ref{unitaritycond}) is fulfilled.
It is immediate to generalize to $S^n{\St}^n=K_n(\varpi){\bf 1}\equiv
N_n^{-1}(\varpi){\bf 1}$ and we get
\beq
(S_0\Sb_0)^n=N_n(\varpi)(S\Sb)^n, \quad N_n(\varpi)=\prod_{l=1}^n
N_1(\varpi+2(l-1))
\label{Nn}
\eeq
 $N_n(\varpi)$ can be connected with the normalizations
$I^{(J)}_m(\varpi)$: Since  we have the relations
$$
{\St}^n=I^{(0)}_n(-\varpi)V_{-n}^{(0)}
$$
\beq
S^{-n}=I^{(0)}_{-n}(\varpi) V^{(0)}_{-n}
\eeq
we can also write
\beq
N_n(\varpi)={I^{(0)}_{-n}(-\varpi)\over I^{(0)}_{n}(\varpi)}
\label{Nn-I-rel}
\eeq
Comparing with Eq. (\ref{Nn}), it is then checked immediately that
$I^{(0)}_{-n}(\varpi)$ is given
by the standard analytic continuation prescription
$\prod_{l=1}^{-n}f(l):=
1/\prod_{l=1}^n f(l-n)$ ; more generally one can show that the same
is true
for $I^{(J)}_m$ with $n=J+m <0$.

{}From the states $|y;f_{\varpi_1}\rangle$ we can now form the
projector onto a fixed eigenvalue $e^{iy}$,
\beq
P_{y}:= {1\over 2\pi }\sum_{f_{\varpi_1}}
|y;f_{\varpi_1}\rangle \langle y;f_{\varpi_1}|
\label{proj}
\eeq
where the sum goes over all Fock states over the ground state
$|\varpi_1
\rangle$. The most important property of $P_y$ is the explicit
presence
of negative screenings, which will turn out not to decouple in the
three point
function when $P_y$ is inserted according to Eq. (\ref{insertion}).

A remark about normalization is in order here. Since we
 consider only zero mode momenta which differ from some given
starting
value $\varpi_i$ by  integers\footnote{(when $2J$ is integer, see
below)}, we adopt
the scalar product
\beq
\langle \varpi' \vert \varpi \rangle =\delta_{
\varpi,\varpi'}
\label{scprod}
\eeq
 appropriate for discrete spectra. With this convention, we have, if
$\varpi_1$ is taken to agree with $\varpi_i$ modulo an even integer,
 $\int_0^{2\pi} dy P_y ={\bf 1}$ in the space $\cup_{l\in {\bf Z}}
V_l$, the union of all Verma modules $V_l$ over the ground states
 $\varpi_i+2l$, with $l$ an arbitrary integer. The projector $P_y$
thus works fine for the truncated theory living on a discrete
spectrum
over some arbitrary starting value $\varpi_i$.
 On the other hand, the full Liouville theory has a continous
spectrum and hence the zero mode volume is infinite. Thus
 in order to compare with the DOZZ result at the particular integer
points where it is possible, we will have
to renormalize our three point function by the infinite volume of the
zero mode integration; we will return to this point later.

\subsection{The three point function}
\label{threepointfn}

We are now in a position to write an explicit expression for the
three-point
matrix element (\ref{matel}) for the case where $\alpha_+$ screenings
are absent.
 Inserting the projector $P_y$ according to
(\ref{insertion}),
we have
\beq
P_y |\varpi_i\rangle = {1\over 2\pi}\sum_{l=-\infty}^\infty
 e^{iy({\varpi_1-\varpi_i
\over 2} -l)}
(S_0\Sb_0)^{l-{\varpi_1-\varpi_i\over 2}}\vert \varpi_i\rangle
\label{proj1}
\eeq
whence
$$
\langle\varpi_f| \exp{[-J\alpha_-\Phi]}(\tau=\sigma=0)
P_y|\varpi_i\rangle =
\sum_{n=0}^\infty {T(\varpi_f)\over T(\varpi_f+2n-2J)}
$$
\beq
\times a_n^{(J)}(\varpi_f) e^{iy(n-s)}
\langle \varpi_f | V^{(J)}_{-J}S^{n}{S_0}^{s-n} {\Vb}^{(J)}_{-J}
{\Sb}^{n}{\Sb_0}^{s-n} |\varpi_i
\rangle .
\eeq
where $s=J-\demi(\varpi_f-\varpi_i)$ is the screening number of Eq.
(\ref{s}) and we have used Eq. (\ref{shift})
and Eq. (\ref{scprod}) to reduce the $l$ sum from Eq. (\ref{proj1})
to a single term.
Writing $(S_0\Sb_0)^n =N_n(\varpi)(S\Sb)^n$ as in the previous
subsection,
the matrix element of the Coulomb gas vertex operators is just
\beq
\langle \varpi_f | V^{(J)}_{-J}S^{n}{S_0}^{s-n}
 {\Vb}^{(J)}_{-J}{\Sb}^{n}
{\Sb_0}^{s-n} |\varpi_i
\rangle =N_{s-n}(\varpi_f+2n-2J)|I^{(J)}_{s-J}
(\varpi_f)|^2
\eeq
according to Eq. (\ref{Vops}). Evaluating
$a_n^{(J)}(\varpi_f)N_{s-n}(\varpi_f+2n-2J)$ explicitly, we thus
obtain
$$
\langle\varpi_f| \exp{[-J\alpha_-\Phi]} P_y|\varpi_i\rangle =
|I^{(J)}_{s-J}(\varpi_f)|^2
\left({\Gamma(1-h/\pi)\over  2\pi }\right)^{2s}
e^{-isy}
$$
$$
\times
\sum_{n=0}^\infty{T(\varpi_f)\over
 T(\varpi_f+2n-2J)}{\Gamma(\varpi_i h/\pi)\over
\Gamma[(\varpi_f+2n-2J)h/\pi)]
 (\varpi_i+1)_{n+2s}}
$$
\beq
\times (-1)^n{2J\choose n}_q
{\lfloor\varpi_i\rfloor_{n+2s}\over \lfloor \varpi_f+1\rfloor_n}
e^{iny}
\label{proj3point}
\eeq
Here ${2J\choose n}_q \equiv {\lfloor 2J\rfloor !\over \lfloor
2J-n\rfloor!
\lfloor n\rfloor!}$ denotes a $q$-binomial, defined for negative
integer
$2J=-p$ by ${2J\choose n}_q ={\prod_{i=0}^{n-1}\lfloor -p-i\rfloor
\over \lfloor n\rfloor!}$. Pochhammer symbols $(a)_n$  for negative
$n$ are given by $(a)_n=(-1)^n/(1-a)_{-n}$, and likewise for their
$q$ - equivalents.
  Eq. (\ref{proj3point}) may be rewritten
$$
\langle\varpi_f| \exp{[-J\alpha_-\Phi]} P_y|\varpi_i\rangle
 =
{T(\varpi_f)\over
T(\varpi_f-2J)}N_s(\varpi_f-2J)|I^{(J)}_{\demi(\varpi_i-\varpi_f)}
(\varpi_f)|^2
$$
\beq
\times e^{-iys}
F^{(q)}(-2J,\varpi_f-2J;\varpi_f+1;z)
\label{hypgeom}
\eeq
with $z:=e^{iy}$. Here, $F^{(q)}(a,b;c;z):=\sum_{n=0}^\infty
{\lfloor a\rfloor_n \lfloor b\rfloor_n \over \lfloor c\rfloor_n
\lfloor n\rfloor!} $ is the q-hypergeometric function \cite{VK}, with
$\lfloor a\rfloor_n:= \lfloor a\rfloor \dots \lfloor a+n-1\rfloor$.
Now $F^{(q)}(a,b;c;z)$ is a priori well-defined only for $\vert
q\vert
<1$ and $\vert z\vert<1$, while here $\vert q\vert=\vert z\vert =1$.
However, in the special case
\beq
2J \in {\bf Z}
\label{2Jcond}
\eeq
the situation is different.
If $2J$ is a nonnegative integer, the hypergeometric function reduces
to a finite sum, corresponding to the fact that in this case, the sum
in Eq. (\ref{Liouexp}) truncates at $n=2J$. We then recover the
Liouville exponentials
for the degenerate (Kac's table) subsector . Though this case is
trivial from the present point of
view, as no resummation is  necessary, we list it for completeness:
$$
\langle \exp{[-J_f\alpha_- \Phi]}\exp{[-J\alpha_-\Phi]}
\exp{[-J_i\alpha_-\Phi]}\rangle^{(J\ge 0)}=
$$
\beq
 g^{x_f}_{J,x_f+m}{\bar g}^{x_f}_{J,x_f+m} {T(-\varpi_0)\over
T(\varpi_0)}
\label{2Jpos}
\eeq
Of course this result could have been read off directly from
Eq. (\ref{Liouexp})
using naive charge conservation rules. We now turn to the interesting
case
\beq
2J =-p
\eeq
with $p$ a positive integer. In this case as well, the hypergeometric
function
extends to $|q|=1$ and $|z|=1$ (in fact to arbitrary $q$ and $z$) via
the transformation formula \cite{VK}
$$
F^{(q)}(p,p+\varpi_f;1+\varpi_f;z)={1\over (q^{p-1}-q^{1-p}z)
(q^{p-3}-q^{3-p}z)\dots
(q^{1-p}-q^{p-1}z)}
$$
\beq
\times F^{(q)}(1-p,1+\varpi_f-p;1+\varpi_f;z)
\label{Ftrafo}
\eeq
where the hypergeometric function on the right-hand side is again a
finite sum.
 For the matrix element Eq. (\ref{hypgeom}), the reflection
symmetry Eq. (\ref{thetaex}), which takes
 $V^{(J)}_{-J}S^n\Vb^{(J)}_{-J}\Sb^n \to
\Vt^{(J)}_{-J}{\St}^n { \Vtb}^{(J)}_{-J}{\Stb}^n,
\quad \varpi \to -\varpi$, is readily translated into the condition
that the right-hand side of Eq. (\ref{proj3point}) must be invariant
under
$\varpi_{f,i} \to -\varpi_{f,i}, \quad y \to -y$. (Notice we demand
reflection invariance of the exponential in the middle separately,
and not
only of the matrix element as a whole, which would be a triviality).
 It is now easy to check that reflection invariance indeed
holds for any integer $2J$ with the expression Eq. (\ref{T}), at any
nonsingular
point $z$.
 To arrive at the three point matrix element
Eq. (\ref{matel}), it remains to perform the $y$ - integration.
Eq. (\ref{Ftrafo}) exhibits first order poles at
\beq
z_j=q^{2p-2-2j}, \quad
j=0,\dots 2(p-1)
\label{poles}
\eeq
 for which we need a prescription. There are three natural
possibilities, $y \to y\pm i\epsilon$ and taking the Cauchy principal
value, and we will discuss all of them. In all cases, the  integral
over
$y$ or $z$ reduces to a sum over residues, and we have
\beq
\langle\varpi_f \vert \exp{[-J\alpha_-\Phi]}
\vert \varpi_i\rangle=
|I^{(J)}_{\demi(\varpi_i-\varpi_f)}(\varpi_f)|^2 K_1(\varpi_f)
K_2(\varpi_i) (R_0+R_1)
\label{matelres}
\eeq
Here,
$$
R_1=\delta{(q-q^{-1})^{-2(p-1)}\over \lfloor 2p-2
\rfloor!}q^{(1-p)(2p-3)}
\sum_{j=0}^{2p-2}q^{(2s+2)(1-p+j)+j(2p-3)}
$$
\beq
\times (-1)^{j+1}
{2p-2\choose j}_q
 F^{(q)}(1-p,1+\varpi_f-p;1+\varpi_f;
q^{2p-2j-2})
\label{ressum}
\eeq
is the sum of the residues at $|z|=1$, with the factor $\delta $
equal to $\demi$ for the Cauchy principal value prescription, $0$ for
$y\to y+i\epsilon$, and $1$ for $y\to y-i\epsilon$.
Furthermore, $R_0$ is the contribution from the residue at $z=0$,
i.e.
the trivial charge-conserving term which appears only for $s\ge 0$.
It takes the simple form
\beq
R_0={\lfloor p\rfloor_{s} \lfloor
 \varpi_f+s+1\rfloor_{p-1}\over \lfloor
s\rfloor ! \lfloor 1+\varpi_f\rfloor_{p-1}}.
\label{R0value}
\eeq
Of course it could have been obtained also directly from Eq.
(\ref{Liouexp}),
 using naive charge conservation rules.

$R_1$ can be largely simplified. Indeed, one easily  derives the
identity
$$
\sum_{j=0}^{2p-2} q^{-j(2n-2s-2p+1)}
(-1)^j {2p-2\choose j}_q=
q^{(1-p)(2n-2s-2p+1)}(q-q^{-1})^{2p-2}
$$
\beq
\times{\lfloor n-s-1\rfloor!
\over \lfloor n-s-2p+1\rfloor!}
\eeq
where  $s=J-\demi(\varpi_f-\varpi_i)$ as above.
Thus we have
\beq
R_1=-\delta{-s-1\choose 2p-2}_q
 \   {_3F}_2^{(q)}(1-p,1+\varpi_f-p,-s;1+\varpi_f,-s-2p+2;1)
\label{Rvalue}
\eeq
Here and below, we define expressions which are
 ill-defined for integer values of $s$ by replacing
$s\to s+\epsilon$ and taking the limit
$\epsilon \to 0$. \footnote{We introduce this convention just
 to avoid clumsy notation; there is
no actual ambiguity in the calculation.}
Eq. (\ref{Rvalue}) can be reduced to a simple product of $q$ -
factorials
again \cite{E}, viz.
\beq
R_1=-\delta{\lfloor 1+s\rfloor_{p-1} \lfloor
 \varpi_f+s+1\rfloor_{p-1}\over \lfloor
p-1\rfloor ! \lfloor 1+\varpi_f\rfloor_{p-1}}.
\label{Rvalue'}
\eeq
Comparing with Eq. (\ref{R0value}), we see that the only
 difference to  $R_0$ is the replacement of  the binomial coefficient
$\lfloor -2J\rfloor_{s}/\lfloor s\rfloor!$ from the expansion
Eq. (\ref{Liouexp})  by $-\lfloor 1+s\rfloor_{-2J-1}/
\lfloor -2J-1\rfloor!$, which can be viewed as an analytic
continuation
of the former. For $s \ge 0$, we have
\beq
\delta \cdot R_0=-R_1
\label{Rrel}
\eeq
One can now check whether
 reflection invariance  holds true for our result Eq.
(\ref{matelres}).
In general,  for the matrix element (\ref{matel}), reflection
invariance
is equivalent to the replacement
\beq
\varpi_f\to -\varpi_f, \, \varpi_i \to -\varpi_i.
\label{refl}
\eeq
We first observe that for the Cauchy principal value prescription
$\delta=
\demi$, reflection invariance is indeed fulfilled, though in a rather
peculiar way:
Under the transformation (\ref{refl}), the  contribution from $R_1$
behaves
{\it anti}symmetrically,
while the one from $R_0$ simply disappears
when we go from positive $s $ to negative ${\tilde s}=-p-s$
(cf. Eq. (\ref{s})). We can express the result very concisely in
terms
of the coupling constants
$g^x_{J,x+m}{\overline g}^x_{J,x+m}$ of Eq. (\ref{gform}).
Indeed, we have
\beq
\langle \varpi_f|\exp{[-J\alpha_-\Phi]}|\varpi_i\rangle=
 \demi \cdot  g^{x_f}_{J,x_f+m}{\bar g}^{x_f}_{J,x_f+m}
{T(\varpi_f)\over T(\varpi_i)}
\label{gg}
\eeq
Eq. (\ref{gg}) needs some
explanation in the case $n=J+m=s<0$, as
the coupling constants corresponding to negative screenings were not
derived
in refs. \cite{CGR1},\cite{GS3}. However, using the associativity of
the
operator product, they can be deduced from the couplings with
$s\ge0$ \cite{JS}; this will be worked out in detail elsewhere.
 We need here only
the result that in the case $s<0$, the coupling constants are still
given
by Eq. (\ref{grel}), with $a_n^{(J)}$ and $I^{(J)}_m(\varpi)$
continued to negative $n=J+m$
as explained below Eq. (\ref{Nn-I-rel}), except that ${\lfloor
-2J\rfloor_n
\over \lfloor n\rfloor!}$ in Eq. (\ref{coeff}) is replaced by
$-{\lfloor
1+n\rfloor_{p-1}\over\lfloor p-1\rfloor!}$. Invoking conformal
invariance
\cite{GN2}, we can promote the result Eq. (\ref{gg}) for the ground
state
 matrix element to an operator equation giving an "effective"
representation for the Liouville exponentials Eq. (\ref{Liouexp}):
\beq
\exp{[-J\alpha_-\Phi]}_{\hbox{{\rm eff}}}=\sum_{n=-\infty}^{\infty}
{T(\varpi)\over
T(\varpi+2m)}
\demi g^x_{J,x+m}{\overline g}^x_{J,x+m}V^{(J)}_m\Vb^{(J)}_m
\label{effrep}
\eeq
for $2J=-p<0$. By definition, Eq. (\ref{effrep}) is to be evaluated
using naive charge conservation, in contrast to Eq. (\ref{Liouexp}).
It is
Eq. (\ref{effrep}) which should be viewed as the true chiral
decomposition
of the Liouville exponentials. Intriguingly, Eq. (\ref{effrep}) can
be reexpressed in terms of the original expansion Eq. (\ref{Liouexp})
 as follows:
$$
\exp{[-J\alpha_-\Phi]}_{\hbox{{\rm eff}}}=\demi\left
(\sum_{n=0}^{\infty} {T(\varpi)\over
T(\varpi+2m)}
 g^x_{J,x+m}{\overline g}^x_{J,x+m}V^{(J)}_m\Vb^{(J)}_m
+ \right.
$$
\beq
\left. \sum_{{\widetilde n}=0}^{\infty} {T({\widetilde\varpi})
\over
T({\widetilde\varpi}+2{\widetilde m})}
 g^{\widetilde x}_{J,{\widetilde x}+{\widetilde m}}{\overline
g}^{\widetilde x}_{J,{\widetilde x}+{\widetilde m}}
\Vt^{(J)}_{\widetilde m}\Vtb^{(J)}_{\widetilde m}\right )
\label{symmform}
\eeq
where ${\widetilde\varpi}=-\varpi$, $\varpi_0+2{\widetilde
x}=-\varpi$,
and the $\Vt$ fields are formed from the free field $\vartheta_2$ as
in
Eq. (\ref{tildeshifts}).
To show this, let us first observe that the sum over negative $n$ in
Eq. (\ref{effrep}) effectively starts only at $n\le 2J$
as the coefficients $g^x_{J_1,x+m_1}{\overline g}^x_{J_1,x+m_1}$
vanish for
$2J_1+1\le n_1\le -1$.
Now the reflection symmetry discussed above tells us that
\beq
{T(\varpi)\over T(\varpi+2m_1)}g^x_{J_1,x+m_1}{\overline
g}^x_{J_1,x+m_1}
={T(-\varpi)\over T(-\varpi-2m_1)}g^{\tilde x}_{J_1,{\tilde
x}-m_1}{\overline g}^{\tilde x}_{J_1,{\tilde x}-m_1}
\label{reflsymm}
\eeq
The shift $-m_1={\widetilde m}_1$ corresponds to
$\tilde n_1=J_1-m_1=2J_1-n_1\ge 0$ and so we recover the coefficients
of the $n_1\ge 0$ part, evaluated at $-\varpi={\widetilde \varpi}$.
Furthermore, we have $V^{(J_1}_{m_1}=\Vt^{(J_1)}_{-m_1}$ and so
indeed
the $n<0$ summation in Eq. (\ref{effrep}) has the same form as the
$n\ge 0$
part with $\vartheta_1$ replaced by $\vartheta_2$.
Thus, the whole effect of the complicated analysis above is to
symmetrize
Eq. (\ref{Liouexp}) in terms of the $\vartheta_1$ and $\vartheta_2$
representations!\footnote{The picture will however be a little bit
more complicated in the two screening case to be discussed below.} Of
course,
this apparent simplicity is somewhat deceiving as the relation
between the two
free fields is very complicated, and it is by no means trivial to
obtain
higher point correlation functions from Eq. (\ref{effrep}) or
Eq. (\ref{symmform}).

Let us now turn to the second possibility for treating the residues.
The $i\epsilon$ -
prescription is consistent with reflection invariance only
 if we employ the $s$-dependent prescription
$y\to y+i\epsilon \, \hbox{sgn} s$, with
$\hbox{sgn} s:= 1$ for $s\ge 0$, and $-1$ otherwise. In this case,
 there is a residue contribution only for $s<0$ and one obtains
$$
\langle\varpi_f|\exp{[-J\alpha_-\Phi]}|\varpi_i\rangle=
$$
\beq
 1 \cdot  g^{x_f}_{J,x_f+m}{\bar g}^{x_f}_{J,x_f+m} {T(\varpi_f)\over
T(\varpi_i)}
\label{iepsilon}
\eeq
and similarly in place of Eq. (\ref{effrep}).
Thus the two choices differ only by an overall normalization factor.
Within the discrete spectrum/finite zero mode volume theory
considered here,
 the proper normalization can be determined by looking at the
operator product expansion of two Liouville exponentials. This will
be done
in the next paragraph, and one finds that the choice Eq. (\ref{gg})
is preferred over Eq. (\ref{iepsilon}).

Eq. (\ref{effrep}) holds for $2J<0$; in the case $2J\ge 0$ where the
sum in
 Eq. (\ref{Liouexp}) is finite and we have only charge-conserving
terms, the
effective representation of course coincides with Eq. (\ref{Liouexp})
itself,
or Eq. (\ref{gform}). We can unify both cases through the formula
Eq. (\ref{symmform}), which trivially holds also when $2J\ge 0$ (cf.
section
 \ref{SL2}).

We can now write down the full three point
function, making use  of the identification
Eq. (\ref{extstates}). For this we still need the
 normalization factors $F_i,F_f$ in Eq. (\ref{extstates}),
which are easily determined from Eqs. (\ref{coeff}), (\ref{T})
and $\lim_{\tau \to -\infty}(z\zb)^{-\Delta_J} V^{(J)}_{-J}\Vb
^{(J)}_{-J}(\sigma,-i\tau)
 \vert \varpi_0\rangle=1\vert \varpi_J \rangle$, with $\varpi_J=
\varpi_0+2J\equiv 1+\pi/h+2J$, and similarly for the left external
state.
 For integer $2J_i,2J_f$ we have
the simple relations
$$
F_i=\epsilon_i{T(\varpi_i)\over T(\varpi_0)}
%=
%{\lfloor \varpi_i\rfloor \over \lfloor \varpi_0
%\rfloor} {\varpi_0!\over \varpi_i!}{\Gamma(1-\varpi_i h/\pi)
%\over \Gamma(1-\varpi_0 h/\pi)}
$$
\beq
F_f= \epsilon_f{T(-\varpi_0)\over T(\varpi_f)}
%=
%{\lfloor -\varpi_0\rfloor \over \lfloor \varpi_f\rfloor}{\varpi_f!
% \over %(-\varpi_0)!}
%{\Gamma(1+\varpi_0 h/\pi)\over \Gamma(1-\varpi_f h/\pi)}
\label{Nif}
\eeq
with $\epsilon_{i,f}=1$ for $2J_{i,f}\ge 0$, and
$\epsilon_{i,f}=\demi$
otherwise, according to Eq. (\ref{effrep}). We would like to stress
here
that there is no restriction on $\varpi_i$, $\varpi_f$ in Eq.
(\ref{gg})
 other than Eq. (\ref{s}), so that up to this point,
$\varpi_f,\varpi_i$
can take continous values. It is only when we try to represent the
states
$|\varpi_{i,f}\rangle$ in terms of Liouville exponentials that we
have
to respect the constraint Eq. (\ref{2Jcond}) arising in our quantum
treatment of the latter.
Combining Eqs. (\ref{Nif}), (\ref{extstates})  and (\ref{gg}), we
obtain
\beq
\langle \exp{[-J_f\alpha_- \Phi]}\exp{[-J\alpha_-\Phi]}
\exp{[-J_i\alpha_-\Phi]}\rangle=
\epsilon_i\epsilon_f\epsilon_J
{T(-\varpi_0)\over T(\varpi_0)}g^x_{J,x+m}{\overline g}^x_{J,x+m}
\label{threepointresult}
\eeq
 We have not yet discussed the dependence
on the cosmological constant; likewise, we did not fix the overall
$c$ - number normalization freedom
\beq
\exp{[-J\alpha_-\Phi]}\to c_J
\exp{[-J\alpha_-\Phi]}
\label{normfreedom}
\eeq
 which is left over by the locality analysis. In refs.
\cite{OW},\cite{GS3}
it was observed that the ambiguity (\ref{normfreedom}) can be fixed
by imposing the leading order multiplicativity property
$\exp{[-J\alpha_-\Phi]}\exp{[-J'\alpha_-\Phi]}\sim
\exp{[-(J+J')\alpha_-
\Phi]}$ and the quantum equations of motion to be $c_J=c^J$, with
\beq
c={\pi\sin h \over 8h^2} \Gamma^2(1+h/\pi)
\eeq
Reinstating the cosmological constant provides another factor
$(\mu^2)^J$, so that the correctly normalized three point function
finally becomes
$$
\langle\exp{[-J_f\alpha_- \Phi]}\exp{[-J\alpha_-\Phi]}
\exp{[-J_i\alpha_-\Phi]}\rangle_{\mu^2}=
$$
\beq
(\mu^2 {\pi\sin h\over 8 h^2}
\Gamma^2(1+h/\pi))^{J+J_i+J_f+\varpi_0}
\langle \exp{[-J_f\alpha_- \Phi]}\exp{[-J\alpha_-\Phi]}
\exp{[-J_i\alpha_-\Phi]}\rangle
\label{norm3point}
\eeq
where we have added  a contribution $\varpi_0$ to the exponent to
account
for the contribution of the background charge to the cosmological
constant dependence \cite{DDK}.

\subsection{Operator product with a degenerate field}
\label{opprod}

Due to the infinite sum in Eq. (\ref{effrep}), the operator product
of two general exponentials is rather delicate. In the present
analysis,
we will restrict ourselves to the case where one of them is
degenerate, which in the single screening case means $2J$ nonnegative
integer. Then the operator
product with an arbitrary conformal field is controlled by the null
vector
decoupling equations \cite{GN2}\cite{G1} and the situation is much
clearer.
Consider therefore
$$
\exp{[-J_1\alpha_-\Phi]}_{\hbox{{\rm eff}}}(\tau,\sigma)
\exp{[-J_2\alpha_-\Phi]}(\tau',\sigma')
$$
with $2J_1=-p<0$ and $2J_2\ge 0$. We take $e^{-J_1\alpha_-\Phi}_{
\hbox{{\rm eff}}}$
to be given by Eq. (\ref{symmform}), except that the factor $\demi$
is
replaced by a general normalization $\delta_{J_1}$ to be determined
(cf. the remark
below Eq. (\ref{iepsilon})), i.e. $e^{-J_1\alpha_-\Phi}_{
\hbox{{\rm eff}}}=
\delta_{J_1}(e^{-J_1\alpha_-\Phi}_{\vartheta_1}+
e^{-J_1\alpha_-\Phi}_{\vartheta_2}).$
For the above operator product, the triangular inequalities
\cite{CGR1}
\beq
-|J_1-J_2|\le J_{12}\le J_1+J_2
\label{triang}
\eeq
are valid , and so only a finite number of Liouville exponentials
$e^{-J_{12}\alpha_-\Phi}$
will appear on the right hand side. As a side remark, we note that
the leading short-distance singularity in the product is given  by
$J_{12}=
J_1+J_2$ as expected if and only if
\beq
J_1-J_2\ge -\varpi_0
\label{Seiberg}
\eeq
For $J_2=-J_1$, this becomes the Seiberg bound \cite{S} for $J_1$! We
leave
the significance of this observation to further study.
Let us now first consider the contribution to
$e^{-J_1\alpha_-\Phi}_{\hbox{{\rm eff}}}$ from the $\vartheta_1$ -
representation. For this part, the fusion was already
written in ref. \cite{GS3} (within the coordinates of the
sphere)\footnote{
We suppress the usual short-distance factor.}:
$$
\exp{[ -J_1\alpha_-\Phi]}_{\vartheta_1}(z_1, \zb_1 )
\exp{[ -J_2\alpha_-\Phi]}(z_2, \zb_2 )=
$$
$$
\sum_{ J_{12}}  \sum _{\{\nu\}, \{\nub\} }
\exp{[ -J_{12} \alpha_-\Phi^{\{\nu\},
\{\nub\}}]}_{\vartheta_1}(z_2,
\zb_2 )\times
$$
\beq
< \!\varpi_{J_{12}}; \{\nu\}, \{\nub\} |
\exp{[ -J_1\alpha_-\Phi]}_{\vartheta_1}
(z_1-z_2, \zb_1-\zb_2 ) |
 \varpi_{J_2} \! >.
\label{fus5}
\eeq
Here, $\{\nu\},\{\nub\}$ denote arbitrary descendants,  and
$\varpi_J=\varpi_0+2J$. Eq. (\ref{fus5}), which was derived from the
chiral
fusion algebra, is
a priori valid only on a formal level as the Liouville exponentials
were treated as formal power series in screening charges in ref.
\cite{GS3}.
Indeed, for general $J_1,J_2$ there is no truncation condition
cutting off
large negative $J_{12}$ so that infinite short-distance singularities
would arise in Eq. (\ref{fus5}). However, if at least one of the
exponentials
is degenerate so that the inequalities (\ref{triang}) hold this
problem is
absent and Eq. (\ref{fus5}) is valid as a consequence of the null
vector decoupling equations.  Let us now turn to the $\vartheta_2$ -
contribution.
 The Liouville exponentials with $2J\ge 0$ are already manifestly
reflection invariant \cite{GS3} (the replacement $\vartheta_1\to
\vartheta_2$ amounts only to a renumbering of terms), and so we can
trivially
substitute $\vartheta_2$ for $\vartheta_1$ in $e^{-J_2\alpha_-\Phi}$.
Then of course Eq. (\ref{fus5}) is again valid for the operator
product,
with the obvious replacements, and the two contributions add up
to form a symmetrized outgoing exponential in the form Eq.
(\ref{symmform})
again\footnote{The exchange
$\vartheta_1\leftrightarrow\vartheta_2$
in the matrix element on the right hand side of Eq. (\ref{fus5})
amounts only to a trivial renaming of variables.} so that we get
$$
\exp{[ -J_1\alpha_-\Phi]}_{\hbox{\rm eff}}(z_1, \zb_1 )
\exp{[ -J_2\alpha_-\Phi]}(z_2, \zb_2 )=
$$
$$
\rho_{J_{12}}\sum_{ J_{12}}  \sum _{\{\nu\}, \{\nub\} }
\exp{[ -J_{12} \alpha_-\Phi^{\{\nu\}, \{\nub\}}]}_{\hbox{
\rm eff}}(z_2, \zb_2 )
\times
$$
\beq
< \!\varpi_{J_{12}}; \{\nu\}, \{\nub\} |
\exp{[ -J_1\alpha_-\Phi]}_{\vartheta_1}(z_1-z_2,
\zb_1-\zb_2 ) |
 \varpi_{J_2} \! >.
\label{fus5'}
\eeq
with $\rho_{J_{12}}=2\delta_{J_1}$ for $J_{12}\ge 0$ where the
Liouville
exponentials are known to have unit normalization,
and $\rho_{J_{12}}=\delta_{J_1}/\delta_{J_{12}}$ for $J_{12}<0$.
  Let us consider the case $J_2=-J_1=-J$ now. To
leading order, we expect the relation
\beq
\exp{[-J\alpha_-\Phi]}(\tau,\sigma)
\exp{[+J\alpha_-\Phi]}(\tau',\sigma')\sim {\bf 1}
\eeq
On the other hand, Eq. (\ref{fus5}) gives
\beq
\exp{[-J\alpha_-\Phi]}(\tau,\sigma)
\exp{[+J\alpha_-\Phi]}(\tau',\sigma')\sim 2\delta_J {\bf 1}
\eeq
and so we conclude that $\delta_J=\demi$ as in Eqs. (\ref{effrep})
and
(\ref{symmform}), and $\rho_{J_{12}}=1$ in all cases.

\smallskip
We close this section with a few remarks on locality.
The locality properties of the Liouville exponentials in the form
Eq. (\ref{Liouexp}) or Eq. (\ref{genexp}) were analyzed in great
detail
in ref. \cite{GS3}. However, one may wonder whether this analysis is
affected
by the new nonperturbative contributions derived above for $J<0$.
We have seen that in the region $s=n\ge 0$, the only effect of the
residue contribution is a renormalization of the coefficients in
Eq. (\ref{Liouexp}) by a factor $\demi$. On the other hand,
the positive and negative powers of  screenings remain of the same
type
under braiding \cite{JS} and so the mutual locality of the positive
screening  parts of the exponentials as analyzed in ref. \cite{GS3}
is
certainly a necessary condition for locality. Reflection invariance
then guarantees that also the negative screening contributions are
mutually
local.
On the other hand, one still has to check the relative locality of
positive
and negative screening contributions; this will be shown elsewhere
within
a detailed analysis of the fusion and braiding symbols for negative
screenings.
At this point, we would like to make contact with ref. \cite{T},
where recursion relations for the
Liouville three point function are derived by exploiting crossing
symmetry for a four point function with one degenerate (spin $\demi$)
exponential. As explained  in ref. \cite{G5}, these are just the same
 recursion relations as those fulfilled by
 the coupling constants $g^x_{J,x+m}{\overline g}^x_{J,x+m}$
of the chiral
operator product algebra (for the latter, they
 were derived in ref. \cite{CGR1})\footnote{(within the special
context
of positive half-integer spins, but they can be immediately extended
- see
ref. \cite{G5}).}. In fact the crossing symmetry conditions
as analyzed in \cite{T} are equivalent to demanding that
$[e^{-\demi\alpha_{\pm}\Phi}(\tau,\sigma),
e^{-J\alpha_-\Phi}(\tau,\sigma')]=0$,
that is, locality between the spin $\demi$ exponential with both
screening
types and any other one with arbitrary spin $J$. As this problem is
completely
controlled by the null vector decoupling equations obeyed by
$e^{-\demi
\alpha_\pm\Phi}$ \cite{GN2}, for this particular situation one
automatically
obtains the exact nonperturbative locality conditions. Remarkably,
 these conditions, although they pertain only to a very
particular case, are stringent enough to determine the most general
three point couplings under the assumption that the latter are
continous in the
spins; this is because the set of shifts $\Delta
J=n+\nhat{\alpha_+\over \alpha_-}$ lies dense in the real numbers
provided $\alpha_+,\alpha_-$
are incommensurate, which is the case for generic values of the
central
charge.  However in the truncated theory considered here, where only
integer $2J$ and $s$ are admitted, this argument works only within
the separate regions $2J\ge 0$ and $2J<0$, or $s>0$ and $s\le 0$, as
the ratio of the coefficients at the boundary is either zero or
ill-defined. As the same functional form of the coefficients
governs both regions $2J\ge 0$ (where locality is known to hold as a
consequence of the null vector decoupling equations
\cite{GN2}\cite{G1}),
and $2J<0$, it is not surprising
that the above crossing symmetry recursion relations are indeed
fulfilled
also by the coefficients in Eq. (\ref{effrep}) in the region $2J<0$;
a more detailed exposition
of this point can be found in ref. \cite{G5}.
\section{The principle of monodromy invariance and the extension
to the two screening case}
\label{monodromy}
To explain the basic idea, let us first consider the classical
theory.
Eq. (\ref{classfield}) gives for general Liouville
exponentials
$$
\exp{[-J\varphi]}=(AA'^{-\demi}B'^{-\demi})^{2J}
(1-{B\over A})^{2J}=
$$
\beq
(AA'^{-\demi}B'^{-\demi})^{2J} \sum_{n=0}^\infty {2J \choose n}
(-1)^n
\left({B\over A}\right)^n.
\label{classexp}
\eeq
$B/A$ is just the classical version of the screening operator $S\Sb$
while
$(AA'^{-\demi}B'^{-\demi})^{2J}$ corresponds to
$V^{(J)}_{-J}\Vb^{(J)}_{-J}$, and
Eq. (\ref{Liouexp}) is the quantum version of
of Eq. (\ref{classexp}).
In the elliptic sector, one has \cite{GN1} $\left | {B\over A}
\right | \equiv 1$.
There is a slight subtlety in the passage from Eq. (\ref{classfield})
to Eq.
(\ref{classexp}) which turns out to have important consequences.
Namely,
Eq. (\ref{classfield}) is single-valued (as it must be, since the
classical
Liouville field is real), while Eq. (\ref{classexp}) is
not\footnote{This
was already noted in ref. \cite{GS3} in the context of the
periodicity properties of the Liouville exponentials.}. Indeed, if
$A$ and $B$ are
reexpressed in terms of chiral free fields \cite{GN1}\cite{LS},
 the dependence
on the classical free field zero mode $q_0$ becomes
$A{A'}^{-\demi}B'^{-\demi}\sim e^{-iq_0}$ and ${B\over A} \sim
e^{2iq_0}$. Then under a rotation $q_0 \to
q_0+2\pi$ around the unit circle, Eq. (\ref{classexp}) picks up a
factor
$e^{2\pi iJ}$ (the binomial series is periodic term by term), while
Eq. (\ref{classfield}),
and  any power of it, is invariant. Consequently, when performing the
zero mode integration in the quantum case within the evaluation of
some matrix element,
we are  not entitled to use the expansion Eq. (\ref{Liouexp})
throughout
the full range $q_0 \in (-\infty,+\infty)$, but only in some interval
of length $2\pi$. A  remedy exactly appropriate for this problem was
proposed in ref. \cite{PRY} within the context of the calculation
of conformal blocks for theories based on $SL(2)$ Kac-Moody current
algebra.
For pedagogical purposes, let us consider instead of Eq.
(\ref{classexp})
the simplified toy version
\beq
f_J(z):= z^{-J}(1-z)^{2J}
\eeq
with $|z|=1$.
The nontrivial monodromy of $f_J(z)$ around the origin captures the
essence
of the non-singlevaluedness properties of Eq. (\ref{classexp}) or Eq.
(\ref{Liouexp}). A three point correlator would then be modeled by
the product
$f_{J_1}(z)f_{J_2}(z)f_{J_3}(z)$. It was observed in \cite{PRY} that
on the unit circle, there is an infinity of equivalent
representations
of $f_J(z)$,
\beq
f_J^{(\gamma)}(z)=z^{J}\sum_{n=-\infty}^\infty {2J \choose n+\gamma}
z^{n+\gamma}(-1)^n e^{i\pi\gamma}
\label{gammaexp}
\eeq
with $\gamma$ arbitrary real, which can be taken to be real and to
agree with $f_J(z)$ within
some fundamental interval for $\arg z$ (say $\arg z \in (-\pi,\pi)$),
while
differing from $f_J(z)$ by their monodromy around $z=0$. But then, if
in
a ``three point function''
$f^{(\gamma_f)}_{J_f}(z)f^{(\gamma)}_{J}(z)
f^{(\gamma_i)}_{J_i}(z)$ we choose $\gamma_i,\gamma,\gamma_f$ such
that
$\gamma_f-J_f+\gamma-J+\gamma_i-J_i$ is integer, the total monodromy
vanishes and  the non-singlevaluedness problem goes away. This means
that the expansion of the observables has to be chosen
according to the correlator resp. matrix element under consideration.
In the true quantum theory we would consider the expression
$\langle \exp{[-J_f\alpha_- \Phi]}\exp{[-J\alpha_-\Phi]}
\exp{[-J_i\alpha_-\Phi]}\rangle$, or equivalently,  (\ref{matel}),
where the monodromy properties of the two "external"  primaries have
been
encoded in  the corresponding
states according to Eq. (\ref{extstates}) through the values of
$\varpi_f,
\varpi_i$.
Using the expansion Eq. (\ref{Liouexp}) for the exponentials
 corresponds to the
case $\gamma_f=\gamma=\gamma_i=0$ above. Indeed, from Eq.
(\ref{Liouexp}) we
infer that
the total dependence on the quantum zero mode $q_0$ - rescaled by a
factor
$\alpha_-$ with respect to the classical one \cite{GS3} - is
$\exp{[q_0
\alpha_-(J_f+J+J_i+\varpi_0+r)]}$, where $r$ is an integer, and
the term $\varpi_0$ counts the difference in the
momenta of the left and right vacuum, which is given by the
background charge.
Thus, monodromy invariance requires
$J_f+J+J_i+\varpi_0=0$
modulo integers, or $\demi(\varpi_f-\varpi_i)-J$ integer.  In order
to obtain more general
three point functions, we should introduce expansions for the
Liouville
exponentials  with nontrivial $\gamma$. However,
 according to Eq. (\ref{extstates}),  the  monodromy parameters of
the exterior states in the matrix element
(\ref{matel}) are always fixed to be $\gamma_i=\gamma_f=0$. This is
not a problem, as the full "monodromy deficit" can be absorbed by
$\gamma$.
Now general $\gamma$ - expansions of the quantum
version of $\left(1-{B\over A}\right)^{2J}$ will involve arbitrary
real powers
$(S\Sb)^{n+\gamma}$ of screenings - defined by taking their zero mode
shifts to be $\Delta \varpi=-2(n+\gamma)$ - corresponding to general
discrete
representations of $U_q(sl(2))$. The algebra of these objects is
unknown
to date, and also we cannot expect the modification of the
coefficients
$a_n^{(J)}(\varpi)$ to be given by the simple replacement $n\to
n+\gamma$ as
in the above toy example; rather it should follow from repeating the
locality
analysis of ref. \cite{GS3} in this more general situation. However,
there
is an important special case where the solution is already known:
When $\gamma_f+\gamma+\gamma_i=\gamma=\nhat {\pi\over h}$, we have
$(S\Sb)^{n+\gamma}=(S\Sb)^{n+\nhat\pi/h} \sim (S\Sb)^n(\Shat\Sbhat)
^\nhat$ as is easily checked by comparing the zero mode shifts. For
this
situation, the locality analysis was already performed in ref.
\cite{GS3},
and the result was precisely Eq. (\ref{genexp}). Thus we are lead to
identify
Eq. (\ref{genexp}) with the  expansion of the Liouville
exponentials carrying  monodromies appropriate for matrix elements
with
\beq
\demi(\varpi_f-\varpi_i)-J=
n+\nhat {\pi\over h}.
\label{monodrcond}
\eeq
with $n,\nhat$ arbitrary integers.
\smallskip
Let us now analyze the three point matrix element (\ref{matel}) in
the general
setting Eq. (\ref{monodrcond}). There are two different points of
view
that we can take, which essentially lead to the same result.
According to the
discussion above, in the situation Eq. (\ref{monodrcond}) where
nontrivial
monodromies are involved, we should actually consider instead of
Eq. (\ref{genexp}) or (\ref{gform}) an expansion of the form
\beq
\exp{[-J^e\alpha_-\Phi]}_{\hbox{{\rm eff}}}=\delta_{J^e}
\sum_{{\underline n}}
{T(\varpi)\over
T(\varpi+2m)}
g^x_{J^e,x+m^e}{\overline
g}^x_{J^e,x+m^e}V^{(J^e)}_{m^e}\Vb^{(J^e)}_{m^e}
\label{gform'}
\eeq
where the $n,\nhat$ sums extend a priori over arbitrary integers.
The values of the coupling
constants outside the region $n,\nhat \ge 0$ would then be determined
directly from locality and we would arrive at an effective
representation
of the form Eq. (\ref{effrep}) without performing a residue
calculation.
Of course, the relative normalizations of regions not connected by
the locality
conditions or reflection invariance cannot be fixed by this analysis.
On the other hand, we can take
Eq. (\ref{genexp}) as a starting point for a nonperturbative analysis
 in the same way as we did for Eq. (\ref{Liouexp}). We will briefly
indicate how the calculation is carried out in this more general
setting.

\subsection{Residue calculation in the two screening case}
The logic of the derivation goes exactly as in section
\ref{threepointfn}.
Instead of Eq.(\ref{proj}), we introduce the projector
\beq
P_{y,\hat y}:=\left({1\over 2\pi}\right)^2\sum_{f_{\varpi_1}}
|y,\hat y;f_{\varpi_1}\rangle \langle y,\hat y;f_{\varpi_1} |
\label{genproj}
\eeq
formed from the eigenstates
\beq
|y,\hat y;f_{\varpi_1}\rangle :=\sum_{n,\nhat =-\infty}^\infty
e^{-iyn}e^{-i\hat y \nhat}
(S_0{\Sb}_0)^n ({\Shat}_0{\Sbhat}_0)^{\nhat}|f_{\varpi_1}\rangle
\eeq
with $\Shat$ as in Eq. (\ref{genexp}).
Starting from the expansion Eq. (\ref{genexp}), all steps in the
above calculation  can then be repeated, if we use that
$[S_0,\Shat_0]=0$ (this follows from $[S,\Shat]=0$, which was shown
in ref. \cite{GS3}). The residue sums factorize into two expressions
of the form Eq. (\ref{ressum}) and Eq. (\ref{R0value}), related by
the exchange $h\leftrightarrow
\hhat\equiv \pi^2/h$.
The restriction
Eq. (\ref{2Jcond}) is replaced by
\beq
2J^e \in {\bf Z}+{\bf Z}{\pi\over h}
\label{2Jgencond}
\eeq
and the screening numbers are subject to Eq. (\ref{gens}). We
slightly
changed the notation $J\to J^e$ to make it clear that $2J$ need not
be integer
any more, and similarly we will write $s^e$ instead of $s$. It will
be convenient to decompose
\beq
J^e=J+\Jhat {\pi\over h}, \quad s^e=s+\shat{\pi\over h}
\eeq
with $s,\shat,2J,2\Jhat$ integer.\footnote{The same notation is used
in refs.
\cite{CGR1},\cite{GS3}.}
The normalization
constants $I_{m^e}^{(J^e)}$ (with $m^e=n+\nhat{\pi\over h}$)
 in this more general situation have been computed in ref. \cite{GS3}
for $n,\nhat$ positive (see appendix) and can immediately be
continued to $n$ or $\nhat$ negative using the argument given in the
previous section.

The result can again be written in a form similar to Eq. (\ref{gg}),
where
residue contributions are relevant. There is a factor $\demi$ if only
one of the spins $J,\Jhat$ is negative so that only
one residue contribution is present, and a factor ${1\over 4}$ in the
case
$J<0,\Jhat <0$ where there are two. The coupling constants are now
given by (cf. \cite{CGR1},\cite{GS3})
\beq
g^{x_f}_{J^e,x_f+m^e}{\bar g}^{x_f}_{J^e,x_f+m^e}=
a_{n,\nhat}^{(J^e)}(\varpi)|I_{m^e}^{(J^e)}(\varpi)|^2
\label{ggendef}
\eeq
for $J,\Jhat\ge 0$. If, say, $J< 0$ but $\Jhat \ge0$ one has to
replace
the factor ${\lfloor -2J\rfloor_n\over\lfloor n\rfloor!}$ in
$a_{n,\nhat}^{(J^e)}$ by
$-{\lfloor 1+s\rfloor_{p-1}\over \lfloor p-1\rfloor!}$ just as below
Eq. (\ref{gg}), and similarly for $J\ge 0, \, \Jhat <0$ and $J<0, \,
\Jhat <0$.
The calculation is well-defined everywhere except for the regions
\begin{eqnarray}
&J\ge 0,\Jhat \ge 0,& \quad s\ge 2J+1 \,\wedge\, \shat \ge
2\Jhat+1\,\vee\, s\le -1\,\wedge\,\shat \le -1 \nnn
&J<0,\Jhat<0,&\quad 2J+1\le s\le -1 \,\wedge\, 2\Jhat+1\le \shat \le
-1
\nnn
&J\ge 0, \Jhat <0,& \quad s\le -1 \,\wedge\, \shat \le -1 \,\vee\,
s\ge 2J+1 \,\wedge\, \shat \ge 2\Jhat+1
\nnn
&J<0, \Jhat \ge 0,& \quad s\le -1 \,\wedge\, \shat\le -1 \,\vee\,
s\ge 2J+1 \,\wedge\, \shat \ge 2\Jhat+1.\nnn
\label{ndefregions}
\end{eqnarray}
For these values of the parameters, poles of the normalizations
$I^{(J^e)}_{m^e}$ occur. However, the full three point function
has poles only for
\beq
J\ge 0,\Jhat<0, \quad s\le -1 \,\wedge\, \shat\le 2\Jhat
\,\vee\, s\ge 2J+1\,\wedge\,\shat\ge0
\label{truepoles}
\eeq
and vice versa,
while in the other cases, the singularities in $|I^{(J^e)}_{m^e}|^2$
 are compensated by a vanishing factor from
the residue calculation or directly from the coefficient
$a^{(J^e)}_{n,\nhat}$
in Eq. (\ref{genexp}), and the product is ill-defined. In these
cases,
the correct result is obtained by taking a limit from continous
$J^e$;\footnote{Recall that the $g^x_{J^e,x+m^e}$, understood as
coupling constants for the chiral algebra, are not subject to the
restriction
Eq. (\ref{2Jgencond}) arising in our treatment of the nonchiral
observables.}
this will be shown in the next paragraph. The parameter range
  Eq. (\ref{truepoles})  does not
couple to the other regions and can thus be suppressed consistently.

\subsection{ Reflection amplitude }
\label{reflectionamp}
Let us start from the trivial observation that the weight formula
for the chiral fields $V^{(J^e)}_m$ as well as the Liouville
exponentials,
\beq
\Delta_{J^e}=-J^e-{h\over \pi}J^e(J^e+1)
\label{weight}
\eeq
has two roots $J^e$ and $J^e_r=-\varpi_0-J^e$.
For the corresponding ground states $|\varpi_{J^e}\rangle \equiv
|\varpi_0
+2J^e\rangle$, the replacement
\beq
J^e\to J^e_r=-\varpi_0-J^e
\label{Jrefl}
\eeq
 amounts to $\varpi_{J^e}\to -\varpi_{J^e}$ and therefore
the reflection invariance $\varpi_f \to -\varpi_f, \, \varpi_i\to
-\varpi_i$
of section \ref{SL2} is generated by combining two such operations.
As any primary conformal field is defined uniquely up to a zero mode
dependent normalization by its conformal weight and its zero mode
shift $m$ \cite{GN2}, we  have the identity\footnote{A related
reasoning was already  used in ref. \cite{G1}, where the symmetry of
the braiding and fusion matrices under the
reflection (\ref{Jrefl}) was observed.}
\beq
V^{(J^e)}_m=V^{(J^e_r)}_m
\label{Vid}
\eeq
for the normalized operators $V_m^{(J^e)}$. On the other hand, the
operation
Eq. (\ref{Jrefl}), which implies
\beq
s^e\equiv s+\shat {\pi\over h}= J^e+m^e \to J^e_r+m^e=
s^e-2J^e-\varpi_0
\eeq
always connects a point in Eq. (\ref{ndefregions}) to a point where
the residue
calculation is well-defined, and vice versa. Thus, for example, the
primaries
appearing in the region $J\ge 0,\Jhat\ge 0,s\ge 2J+1,\shat\ge
2\Jhat+1$
contained in Eq. (\ref{ndefregions}) are the same as those for
$J<0,\Jhat<0,
s\ge 0,\shat \ge 0$. But in the latter region, the analysis of
ref. \cite{GS3} tells us that locality fixes the $g\bar g$
coefficients
up to the common similarity transformation $T(\varpi)$ already
discussed,
and a $J^e$ - dependent c-number normalization constant. This means
that the coefficients of the extension $s\ge 2J+1, \shat\ge 2\Jhat+1$
of
the $J\ge 0,\Jhat\ge 0$  exponentials must agree up to a constant
with those
for $J<0,\Jhat<0,s\ge 0, \shat\ge 0$ which are known; similar remarks
apply to
the other cases. Invoking furthermore reflection invariance as in
Eq. (\ref{reflsymm}), which connects pairwise subcases with given
$J,\Jhat$,
one sees that there is only one $J^e$ - dependent constant that can
be
introduced. Thus we arrive at the conclusion that
\beq
\exp{[-J^e_r\alpha_-\Phi]}=A_R(J^e)\exp{[-J^e\alpha_-\Phi]}
\label{Liourefl}
\eeq
for a suitable constant $A_R(J^e)$, which we will call reflection
amplitude,
in the same sense as in ref. \cite{ZZ}. One may now verify by tedious
explicit
calculations
 that Eq. (\ref{Liourefl}) is fulfilled with the definition of the
coefficients
in the region Eq. (\ref{ndefregions}) by limit from continous $J^e$
as proposed
 above. However, there is a much simpler argument allowing to
determine
the explicit value of $A_R(J^e)$:
Consider the joint  reflection operation
$J_f^e\to (J^e_f)_r, \, J_i^e \to (J^e_i)_r$ , or $\varpi_f\to
-\varpi_f,
\, \varpi_i\to -\varpi_i$, which leaves the three point {\it matrix
element}
(\ref{matel}) invariant. Its only effect on the three point function
therefore comes from the normalization factors $F_i,F_f$ in
Eq. (\ref{extstates}) and the cosmological constant dependence (cf.
Eq. (\ref{norm3point})), and we have
$$
A_R(J^e_f)A_R(J^e_i)={ \langle
\exp{[-{ (J_f)_r}\alpha_- \Phi]}\exp{[
-{ J_r}\alpha_-\Phi]}
\exp{[-{ (J_i)_r}\alpha_-\Phi]}\rangle
\over
\langle \exp{[-J_f\alpha_- \Phi]}\exp{[-J\alpha_-\Phi]}
\exp{[-J_i\alpha_-\Phi]}\rangle}
=
$$
\beq
(\mu^2 {\pi\sin h\over 8
h^2}\Gamma^2(1+h/\pi))^{\varpi_f-\varpi_i}{F_f(-\varpi_f)
F_i(-\varpi_i)\over F_f(\varpi_f)F_i(\varpi_i)}
\eeq
from which we read off, using the permutation symmetry of the three
point function and Eqs. (\ref{Nif}),(\ref{T}),
 that\footnote{up to a sign; we will show
below that the choice in Eq. (\ref{reflamp}) is the right one.}
$$
{A_R(J^e)\over (\mu^2 {\pi\sin h\over 8 h^2}
\Gamma^2(1+{h\over\pi}))^{-\varpi_0-2J^e}}=
{F_f(\varpi_0+2J^e)\over
F_f(-\varpi_0-2J^e)}={F_i(-\varpi_0-2J^e)\over
F_i(\varpi_0+2J^e)}=
$$
\beq
-\epsilon_{J^e}
 {\Gamma(2+{\pi\over h}+2J^e)
\Gamma(2+{h\over\pi}+2\Jehat)\over
\Gamma(-2J^e-{\pi\over h})\Gamma(-2\Jehat-{h\over\pi})}
\label{reflamp}
\eeq
The factor $\epsilon_{J^e}$ is equal to $1$  when $J\ge 0, \Jhat <0$
or vice
versa, where the factors $\demi$ arising from the residue
contributions
cancel; for $J,\Jhat \ge 0$ or $J,\Jhat <0$ it takes the values
 ${1\over 4}$ and $4$, respectively. Eq. (\ref{reflamp}) agrees with
the reflection amplitude proposed in ref. \cite{ZZ}\footnote{except
for
 the factor
$\epsilon_{J^e}$, cf. section \ref{DOZZ}.}; the comparison
with the DOZZ conjecture will be discussed in more detail in section
\ref{DOZZ}.
The equality of the reflection amplitude to the ratio of the
normalization
factors $F_f$ or $F_i$ appearing in Eq. (\ref{extstates}) has an
important consequence: It implies that the matrix element
(\ref{matel}), where $F_{f,i}$
are divided out, is invariant not only under the joint reflection
$\varpi_f \to -\varpi_f, \varpi_i\to -\varpi_i$ considered in section
\ref{SL2}, but under
$\varpi_f \to -\varpi_f$ and $\varpi_i \to -\varpi_i$
separately. Consideration of the case $\varpi_f=0$ or $\varpi_i=0$
shows that we have symmetric and not antisymmetric behaviour, and
thus the sign
choice in Eq. (\ref{reflamp}) is correct. This leads us to identify
the ground states with opposite values of the zero mode:
\beq
|\varpi\rangle \cong |-\varpi\rangle
\label{ident}
\eeq
Such an identification is of course expected since the map from the
Liouville field to the free field is two to one, and the ground
states
$|\varpi\rangle$ and $|-\varpi\rangle$ have the same conformal
weight.
However, as we have seen, its validity requires rather nontrivial
properties of the Liouville exponentials and therefore could not yet
be established in previous  analyses. Our argument is based entirely
on the three point function, which however should be sufficient in a
conformal theory \cite{BPZ}\footnote
{Actually there is the subtle matter of factorization for the
Liouville four point function \cite{P}\cite{S}; however, here by
definition
there is no such problem because we factorize over elliptic states -
cf.
section \ref{conc}.}

At this point, a general remark about the extensions
Eq. (\ref{ndefregions})
is in order. While Eq. (\ref{Liourefl}) is valid only when the latter
are taken into account, we know that, for instance, the unextended
exponentials with $J\ge 0,\Jhat\ge 0$ are perfectly consistent with
locality as they are degenerate fields. It is thus the full
realization
of reflection invariance which requires the extensions. A very
similar
remark applies to the  (two-screened) exponentials with $\Jhat =0$
(or $J=0$),
 which naively reduce
to the single-screening expression Eq. (\ref{Liouexp}) but, according
to Eq. (\ref{ndefregions}), actually
retain summations with nonzero $\nhat$ (or $n$).\footnote{Actually,
it is slightly more
appropriate to speak of $n,\nhat$ rather than $s,\shat$ in Eq.
(\ref{ndefregions}) as these values are not related to residue
calculations
but should be included in the initial sum Eq. (\ref{gform}); however,
thinking in terms of effective representations of the form
 Eq. (\ref{gform'}), the difference is of
little relevance.}

Summarizing, we conclude that
the expansion Eq. (\ref{gform}) should be extended to
the regions Eq. (\ref{ndefregions}) without Eq. (\ref{truepoles}) by
taking limits from continous $J^e$ or making use of Eq.
(\ref{Liourefl}),
and this defines the  summation ranges in Eq. (\ref{gform'}).
According to the remark above Eq. (\ref{ggendef}), we have
$\delta_{J^e}=1$ for $J\ge 0,\Jhat \ge 0$ , $\delta_{J^e}={1\over 4}$
for $J<0,\Jhat<0$, and $\delta_{J^e}=\demi$ otherwise.

We close this section with a remark
on  the question of the $SL(2)$ invariant vacuum mentioned in the
footnote
on Eq. (\ref{extstates}).
The prescription (\ref{extstates}) appears to break reflection
invariance
by its preference of the $\vartheta_1$ representation with respect to
the
vacuum $|\varpi_0\rangle$.  On the other hand, the above extensions
of
the Liouville sums of Eq. (\ref{genexp})
tell us that even in the case $\Jhat=0$, the shifts
$m^e=-J^e$ and $m^e=-J^e_r$ always appear together in any given
exponential.
The corresponding operators have the same $z\to 0$ behaviour on the
vacuum
$|\varpi_0\rangle$,  and thus actually two highest weight states
$|\varpi_f\rangle,|-\varpi_f\rangle$ and
$|\varpi_i\rangle,|-\varpi_i\rangle$
are created instead of just one as supposed in Eq. (\ref{extstates}).
It is easy to check that their coefficients are equal, so that no
breaking
of reflection invariance actually occurs! As the two highest weight
states
can be identified according to Eq. (\ref{ident}), the net effect of
the
"forgotten" contribution for the three point function
is just a global factor of $2$ independent of all
spins. Moreover, one can verify that
using the reflected left and right vacua $|\mp\varpi_0\rangle$
instead of
$|\pm\varpi_0\rangle$ leads to the same result, as a consequence of
the
reflection symmetry of the exponentials. This can be taken as a hint
that,
although there does not
truly exist an $SL(2)$ - invariant vacuum in the Hilbert space
 because on each of the candidate vacua $|\pm\varpi_0\rangle$ one of
the free fields becomes singular \cite{LS}, the physical observables
may be shielded  from this problem by reflection invariance and
behave
as if both vacua $|\pm\varpi_0\rangle$ were truly $SL(2)$ - invariant
\footnote{This would fit in well with the ideas in \cite{T1} where it
is proposed
to formulate the theory on states corresponding to Kac's zeroes in
terms
of Verma rather than Fock modules.}.
We hope to discuss this question in greater detail elsewhere.

\section{Comparison with DOZZ}
\label{DOZZ}

We are now in a position to compare our results with the DOZZ
conjecture.
Our discussion will be brief since a detailed comparison between the
coupling constants $g^x_{J,x+m}{\bar g}^x_{J,x+m}$ describing the
three point function and the DOZZ result has been carried out in ref.
\cite{G5}.
The DOZZ result in the form
of ref. \cite{ZZ} reads
$$
\langle\exp{[-J_f\alpha_- \Phi]}\exp{[-J\alpha_-\Phi]}
\exp{[-J_i\alpha_-\Phi]}\rangle_{\mu^2}=
\left[\pi\mu\gamma(b^2)b^{2-2b^2}
\right]^{(Q-\sum_{k=1}^3\alpha_k)/b}
$$
\beq
\times
{\Upsilon_0\Upsilon(2\alpha_1)
\Upsilon(2\alpha_2)\Upsilon(2\alpha_3)
\over
\Upsilon(\alpha_1+\alpha_2+\alpha_3-Q)
\Upsilon(\alpha_1+\alpha_2-\alpha_3)\Upsilon
(\alpha_2+\alpha_3-\alpha_1)
\Upsilon(\alpha_3+\alpha_1-\alpha_2)}
\label{ZZ}
\eeq
Here, we have
\beq
b=\sqrt{h\over\pi}, \, \alpha_k=-bJ_k, \, Q=b+{1\over b}=
b\varpi_0
\eeq
The symbol $\Upsilon$ denotes a new special function for which
integral
representations were given in \cite{OW}\cite{ZZ}. It fulfills the
functional
relations
$$
\Upsilon(x+b)=\gamma(bx)b^{1-2bx}\Upsilon(x)
$$
\beq
\Upsilon(x+1/b)=\gamma(x/b)b^{2x/b-1}\Upsilon(x)
\label{recrel}
\eeq
with $\gamma(x):={\Gamma(x)\over \Gamma(1-x)}$, as well as
\beq
\Upsilon(x)=\Upsilon(Q-x), \quad \Upsilon(Q/2)=1.
\eeq
Finally, $\Upsilon_0=\left.{d\Upsilon(x)\over dx}\right|_{x=0}$.
The $\Upsilon$ function, and thus the
formula Eq. (\ref{ZZ}), is explicitly symmetric in $\alpha_+$ and
$\alpha_-$, or $h$ and $\hhat={\pi^2\over h}$, and thus  supports our
interpretation of the symmetric expansion
Eq. (\ref{genexp}) as a different representation of the Liouville
exponential rather than  a new observable.
The recursion relations Eq. (\ref{recrel}) were shown in ref.
\cite{T}
to follow from crossing symmetry, as mentioned at the end of section
\ref{nonpert}.
The pole and zero structure of the expression (\ref{ZZ}) is
determined
in terms of the zeroes of the  function $\Upsilon(x)$, which lie at
\beq
x=-b(n+\nhat{\pi\over h})
\label{Upszeroes}
\eeq
with $n,\nhat$ both nonnegative or both negative. Let us consider
Eq. (\ref{ZZ}) as a function of the screening charge $s^e=\varpi_0
+\sum_k J^e_k=-{1\over b}(\sum_k \alpha_k -Q)$. For greater
simplicity
of comparison, we consider generic continous values of $J^e_f,J^e_i$
(subject to condition (\ref{gens})) though we can be sure of the
validity
of relations (\ref{extstates}) only if $J_f,J_i$ also fulfill
Eq. (\ref{2Jgencond}).\footnote{Recall that our result for the matrix
element
(\ref{matel}) is truly valid for continous $\varpi_f,\varpi_i$.}
 Then the position of poles and zeroes in Eq. (\ref{ZZ}) as a
function
of $s^e$ does not depend on $J_f,J_i$; as mentioned in the
introduction,
Eq. (\ref{ZZ}) exhibits poles in $s^e$ at
\beq
s\ge 0, \shat \ge 0 \, \vee\,  s<0,\shat <0 \, \vee\,
 s\le 2J, \shat
\le 2\Jhat\,\vee\,  s\ge 2J+1,\shat \ge 2\Jhat+1
\label{polecond}
\eeq
due to the factors
$\Upsilon(\sum_k\alpha_k-Q)\equiv \Upsilon(-bs)$ and
$\Upsilon(\alpha_3+\alpha_1-\alpha_2)\equiv \Upsilon(-b(2J-s))$ in
the denominator, which however may be (partially) compensated by a
vanishing factor
$\Upsilon(2\alpha_2)$ in the numerator when $J\ge 0,\Jhat \ge 0$ or
$J<0,\Jhat <0$. In these cases,  according to ref. \cite{G5} one
 should define the ratios
$\Upsilon(2\alpha_2)/\Upsilon(\sum_k\alpha_k-Q)$
and $ \Upsilon(2\alpha_2)/\Upsilon(\alpha_3+\alpha_1-\alpha_2)$ for
$s\ge 0,
\shat \ge 0$ or $s<0,\shat<0$
and $s\le 2J,\shat \le 2\Jhat$ or $s\ge 2J+1,\shat\ge 2\Jhat+1$,
respectively,  by formal cancellation of the singular factors
$\Upsilon(0)$ remaining in numerator and denominator after using the
recurrence relations.
Now our result Eq. (\ref{threepointresult}), on the other hand, is
finite at all of the points Eq. (\ref{polecond}).
The difference is due to the fact that, as remarked below Eq.
(\ref{scprod}), we have been working with a discrete zero mode scalar
product. We should thus compare Eq. (\ref{norm3point}) with the
residues
of the DOZZ three point function at these poles, which fulfill the
same crossing symmetry recursion relations as for the nonsingular
points. In doing this, we employed a very useful
observation made in ref. \cite{G5}; there it is shown
that  expression (\ref{ZZ})  is the interpolating function
extending the coupling constants in Eq. (\ref{gg}) to continous
spins. More
precisely, the DOZZ three point function is a meromorphic function
whose
 residues at the poles in $s^e$  agree with the values of
the coupling constants $g\bar g$. We thus find that at the above
points,
Eqs. (\ref{threepointresult}) and (\ref{ZZ}) agree up to an overall
constant
independent of all the spins, except for the factors $\demi$,
${1\over 4}$
arising from the Cauchy principal value prescription for the residue
contributions.  This seems to imply that for the continous spectrum
theory, the
$i\epsilon$  prescription for the treatment of the $z$ - poles
Eq. (\ref{poles}) is favoured, which leads to a unit normalization of
the exponentials
with $J< 0$ and/or $\Jhat<0$.
Consequently, the leading order operator product, which was used to
determine
 the latter, should also differ by a factor
of two resp. four from the one in the continous theory, in the
relevant cases. This may seem surprising, but we must keep in mind
that the system
of intermediate states defining the operator product is rather
different in the two theories. In fact, as suggested in \cite{P}
\cite{S} and supported by
the analysis of \cite{ZZ}, in the continous theory the system of
intermediate
states should actually be parametrized by continous imaginary values
of $\varpi$,
the socalled hyperbolic states  (corresponding to complex spins
$J=-\varpi_0/2+ip$),
while in our analysis we have discrete real values of $\varpi$.
The reason why the discrepancy consists just in a constant factor
depending
only on the sign of $J,\Jhat$ is that the crossing symmetry relations
obtained
 from the null vector decoupling equations  are the same for the
continous and the discrete theory, and thus allow for a change of
normalization in the latter
only when traversing the points $J=0$ resp. $\Jhat=0$ where they
become ill-defined.

It remains to analyze the points
\begin{eqnarray}
&J\ge 0,\Jhat \ge 0,&\quad  0\le s\le 2J\, \wedge\,( \shat
\ge 2\Jhat
+1 \vee \shat \le -1) \,\, \hbox{and}\,h\leftrightarrow \hhat
\nnn
&J<0,\Jhat<0,&\quad 2J+1\le s\le -1 \,\wedge\, (\shat\le 2\Jhat
\,\vee\, \shat
\ge 0) \,\,\hbox{and}\, h\leftrightarrow \hhat
\nnn
&J\ge 0,\Jhat<0,& \quad s\ge 2J+1 \,\wedge\, \shat\le 2\Jhat
\,\vee\,
s\le -1  \wedge \shat \ge 0
\nnn
&\phantom{J\ge 0,\Jhat<0,}& \quad \vee\, 0\le s\le 2J
\,\wedge\,
2\Jhat+1
\le \shat\le -1
\nnn
&\hbox{ and}\, h\leftrightarrow \hhat&
\label{zeroes}
\end{eqnarray}
where the DOZZ formula
predicts  finite nonvanishing values. The cases with
$J\ge 0,\Jhat \ge 0$ and $J<0,\Jhat<0$ are connected by the
reflection
operation, and the same applies to the cases with $J\ge 0,\Jhat<0$
and $J<0,\Jhat\ge 0$.
Looking at Eqs. (\ref{Rvalue'}), (\ref{R0value}) or their
"hatted" counterparts, we see that in these cases one of the residue
sums
$R_0+R_1, {\widehat R}_0+{\widehat R}_1$ vanishes
 while the normalizations $I^{(J^e)}_{m^e}(\varpi)$ are finite, so
that our three point function vanishes.
This is perfectly
consistent with the structure of the chiral coupling constants
$g^x_{J,x+m},
{\bar g}^x_{J,x+m}$ in this region \cite{JS}. Thus, the discrete
theory
constructed here lives only on the poles of the DOZZ three point
function.
The above Liouville exponentials should be viewed
as the direct generalization of those for $2J$ positive integer
considered in the older
works \cite{GN}, where the spectrum was also assumed to be discrete.
However, the resulting conformal field theory is obviously
a truncation of the full Liouville theory which has a continous
spectrum.
To derive the nonvanishing finite values of the DOZZ three point
function at the points Eq. (\ref{zeroes}) from our approach,
we thus would need a
regularization prescription to balance the infinite zero mode volume
against the vanishing of the residue sum; this is the well-known
problem
of coupling constant renormalization, which appears in a different
guise
also in the Goulian/Li approach \cite{GL}\cite{D}\cite{DK}.
At present, we cannot derive the proper regularization procedure from
first principles. This is because any systematic regularization
of the zero mode integration seems to require going away at least
infinitesimally from the integer screening points Eq. (\ref{gens}).
We would then need expansions of the Liouville exponentials with a
general
continous monodromy, which, as explained in section \ref{monodromy},
is a nontrivial problem. However, it is of course possible to
formally
 reproduce
the DOZZ three point function at the above points as well in the
operator
approach. Indeed, the main technical ingredients of the latter are
the algebra
of the chiral operators $V^{(J)}_m$ and the locality condition. We
could
therefore start with an ansatz of the form Eq. (\ref{gform'}) and try
to
determine the coefficients such that locality is fulfilled in the
region
Eq. (\ref{zeroes}). The solution is easy to obtain by a formal
renormalization
of the coefficients $g^x_{J,x+m}{\overline g}^x_{J,x+m}$ above.
Indeed,
the expressions $R_0$, $R_1$ (and their hatted counterparts) where
the zeroes
occur, are given by finite products with one vanishing factor. Taking
out
this factor produces a solution of the crossing symmetry conditions
as written in ref. \cite{T}\footnote{This is not very surprising as
the crossing symmetry conditions involve only ratios of
coefficients.}, and
one obtains agreement with DOZZ again.
{}From this point of view, the problem becomes a matter of imposing
boundary
conditions on the solutions of the locality, or crossing symmetry,
conditions;
in other words, we need to decide on which set of points the solution
should
be vanishing resp. finite. We would like to comment here on an older
result
by Gervais for the three point function of minimal matter coupled to
gravity \cite{G1}. There  dressing operators were introduced which
are
local and have the correct conformal dimension, but looked very
different
from the Liouville exponentials that were expected to play this role,
and
so their meaning was not obvious. In particular, they are not
singlets
under the quantum group\footnote{Already classically, the invariance
under $A\to A+c, \ B\to B+c$ (cf. Eq. (\ref{sl2symm}) ) is broken.}.
{}From the present analysis, it is now clear
that they should be viewed as effective representations of the
Liouville
exponentials for the infinite volume case, at integer "off-shell"
values of the spins away from the poles of the DOZZ three point
function; in other words,
they correspond to a restriction of Eq. (\ref{gform'}) with
renormalized
coupling constants to a subset of Eq. (\ref{zeroes}). Explicitly,
the local operators of \cite{G1} can be written as the fusion
 of a standard exponential
of the form Eq. (\ref{Liouexp}) with screening charge $\alpha_+$ and
$2\Jhat$ positive integer, and a new local
field involving $\alpha_-$, or vice versa. The latter can be written
in the form
\beq
\exp{[-J\alpha_-{\breve\Phi}]}=
C_J\sum_{n=2J+1}^{-1}{T(\varpi)\over
T(\varpi+2m)}{\breve g}^x_{J,x+m}
{\breve {\overline g}}^x_{J,x+m}
V^{(J)}_m\Vb^{(J)}_m
\label{Gervaisexp}
\eeq
with $2J$ a negative integer, and $n=J+m$. We see therefore that we
are in the
situation $J<0,\Jhat\ge 0, 2J+1\le s\le -1 \,\wedge\,
0\le \shat\le 2\Jhat$ contained in Eq. (\ref{zeroes}).

Moreover, one
can show
easily that the coefficients ${\breve g}^x_{J,x+m}
{\breve {\overline
g}}^x_{J,x+m}$ as given in ref. \cite{G1} agree precisely with the
renormalized
coupling constants defined above. Thus we can identify
\beq
\exp{[-J\alpha_-{\breve\Phi}]}\equiv
\exp{[-J\alpha_-\Phi]}_{\hbox{ren}}
\eeq
as claimed, and similarly for the fusion with
$e^{-\Jhat\alpha_+\Phi}$.
 We note that the renormalized exponentials are  to be
treated using the  finite volume zero mode scalar product, as the
infinite
volume already went into the renormalization of the coupling
constants.

\section{Conclusions/Outlook}
\label{conc}

Two main lessons are to be learnt from the above analysis: First,
there is a consistent truncation of Liouville theory that involves
integer powers of screenings, or highest/lowest weight
representations
of $U_q(sl(2))$, only. It is for this truncated theory that the
expansions
of ref. \cite{GS3} can immediately be used; however, their evaluation
turned
out to be rather delicate and required a nonperturbative definition
of the
infinite chiral decompositions representing the observables. The main
principles ensuring that
this problem has a well-defined solution were seen to be locality
and reflection invariance; the latter leads
to a symmetric presence  of  highest and lowest weight
representations,
or both  of the two equivalent free fields.
The general message from this
is that while Coulomb gas type representations similar to the ones
familiar
from rational conformal field theories continue to work for the
chiral
operator algebra in the present irrational theory, the appearance
 of an infinite number of primaries in  the
nonchiral  observables   leads to new nonperturbative effects which
break charge conservation. Within the family of discrete spectrum
truncations of Liouville theory discussed here, this breaking
is not arbitrary but can be parametrized in terms of negative
screening powers.  Remarkably,  the Gervais-Neveu quantization scheme
provides us from the beginning with the appropriate technical
machinery to describe the resulting
generalized Coulomb gas picture and gives an explicit operator
representation of positive and negative screenings on the same
footing; this
  should, in particular,  make further progress
possible for $N\ge 4$ point functions by exploiting the conformal and
group-theoretic properties of this representation.

Second, in order to obtain the three point function in the most
general situation, according to the principle of monodromy invariance
one has to study the algebra
of arbitrary powers of screenings, or general discrete
representations
of $U_q(sl(2))$.  Technically, what is required for
the general case is an extension of the
$q-6j$ symbols to  noninteger screening numbers in such a way
that the polynomial equations of Moore and Seiberg \cite{MS} remain
valid, together
with a Coulomb gas picture for arbitrary screening powers along the
lines of
\cite{PRY}. The final result should be a decomposition formula
for the Liouville
exponentials in terms of arbitrary representations.  This would be
very important in particular in order to resolve the problem of a
rigorous treatment of the zero mode integration in the continous
spectrum
 case, and to address the issue of factorization.
Indeed, from general arguments \cite{S}\cite{P} one expects the four
point function to factorize over hyperbolic states - corresponding to
imaginary
$\varpi$ - and this is supported by the analysis of \cite{ZZ}. On the
other
hand, in the present analysis we have factorized over elliptic states
($\varpi$ real) by means of the coherent state projector. It may well
be that
this is truly consistent only in the truncated theory; in any case,
the meaning of the poles contributing to our residue integral needs
to be better understood. Finally, the issue of
the Seiberg bound still awaits its resolution; neither in the
DOZZ analysis nor in the present one has it been seen to play any
significant role so  far. However, it may  surface once we analyze
the operator
product of the Liouville exponentials - cf. the remark in section
\ref{opprod} - which is expected to have a rather nontrivial
structure in the general case. We leave these interesting
 questions to further studies.

\vskip 5mm
\noindent
{\bf Acknowledgements}
It is a pleasure to thank Jean-Loup Gervais for the fruitful
collaboration
on the earlier work and useful discussions and criticisms on the
present one.
I am also grateful   for   helpful discussions to L. Alvarez-Gaume
and S. Shatashvili.

\section{Appendix }
Here we give the explicit formulae for the normalizations
$I_m^{(J)}(\varpi)$
which were derived in ref. \cite{GS3}. In the case $m=n-J$,
$n=0,1,2,...$,
we have
$$
I_m^{(J)}(\varpi)=(2\pi\Gamma(1+{h\over\pi}))^ne^{ihn(\varpi-J+m)}
$$
\beq
\times\prod_{l=1}^n{\Gamma(1+(2J-l+1)h/\pi)\over
\Gamma(1+lh/\pi)\Gamma(1-(\varpi+
2m-l)h/\pi)\Gamma(1+(\varpi+l)h/\pi)}
\label{norm}
\eeq
for any continous value of $J$. For $n<0$, Eq. (\ref{norm}) is still
valid
if we define as usual
$\prod_{l=1}^{-|n|}f(l):=1/\prod_{l=1}^{|n|}f(l-|n|)$.
If both screening charges are present, so that $m=-J+n+\nhat\pi/h$,
one obtains
\[
I_{m\mhat}^{(J\Jhat)}(\varpi)=I^{(\Je)}_{\ms}(\varpi+2\nhat{\pi\over
h})
{\widehat I}^{(\Jehat)}_{\mshat}
(\varpihat +2n{h\over\pi}) (i{\pi\over h})^{2n\nhat}\times
\]
\[
\prod_{l,\hat l =1}^{n,\nhat}\{ (l+\hat l {\pi\over h})
(\varpi +2m +2\mhat{\pi\over h}
-l-\hat l{\pi\over h})(\varpi+l+\hat l {\pi\over h})
(2\Je -(l-1)-(\hat l -1){\pi\over h})\}^{-1}
\]
\beq
\times\prod_{\hat l=1}^{2\nhat}\prod_{l=1}^n (\hat l +(\varpi
+l){h\over\pi})
\prod_{ l=1}^{2n}\prod_{\hat l=1}^\nhat ( l +(\varpihat +\hat
l){\pi\over h})
\label{gennorm}
\eeq
for nonnegative $n,\nhat$. The continuation to negative $n$ or
$\nhat$
 works in the same way as for Eq. (\ref{norm}).

\end{document}